# Defending against Phishing Attacks: Taxonomy of Methods, Current Issues and Future Directions

[1]B. B. Gupta*, [2]Nalin A.G. Arachchilage, [3]Konstantinos E. Psannis

[1]Department of Computer Engineering, National Institute of Technology Kurukshetra, India
[2]Australian Centre for Cyber Security (ACCS), The University of New South Wales
Australian Defence Force Academy, PO Box 7916, Canberra BC ACT 2610, Australia
[3]School of Information Sciences, University of Macedonia, 54006 Thessaloniki, Greece
*gupta.brij@gmail.com[1], nalin.asanka@adfa.edu.au[2], kpsannis@uom.gr[3]

***Abstract*** **–** Internet technology is so pervasive today, for example, from online social networking to online banking, it has made people's lives more comfortable. Due the growth of Internet technology, security threats to systems and networks are relentlessly inventive. One such a serious threat is "phishing", in which, attackers attempt to steal the user's credentials using fake emails or websites or both. It is true that both industry and academia are working hard to develop solutions to combat against phishing threats. It is therefore very important that organisations to pay attention to end-user awareness in phishing threat prevention.

Therefore, aim of our paper is twofold. First, we will discuss the history of phishing attacks and the attackers' motivation in details. Then, we will provide taxonomy of various types of phishing attacks. Second, we will provide taxonomy of various solutions proposed in literature to protect users from phishing based on the attacks identified in our taxonomy. We conclude our paper discussing various issues and challenges that still exist in the literature, which are important to fight against with phishing threats.

# I. INTRODUCTION

Since the past, one of the most profitable crimes is 'identity theft' [1]. Identity theft is the crime in which criminals steal personal identity or financial information such as banking details [56]. In traditional way as discussed in [2], criminals commit crimes either by killing the victim and pretend to be the legitimate person or steal confidential information from garbage, where criminals access information from discarded letters, financial records, electricity bills, and many others bills which are dumped without shredding properly.

The concept of 'phishing' came from traditional 'fishing', in which a fisher trolls in a boat on the river and uses bait to catch the fish. Similarly, 'phisher' also trolls the Internet by using any communication method and uses bait to convince the user and steal user's credentials. The information provided by phishers seems to be legitimate at first glance. Figure 1, which depicts the exponential growth in the number of Internet users, year by year [41, 42].

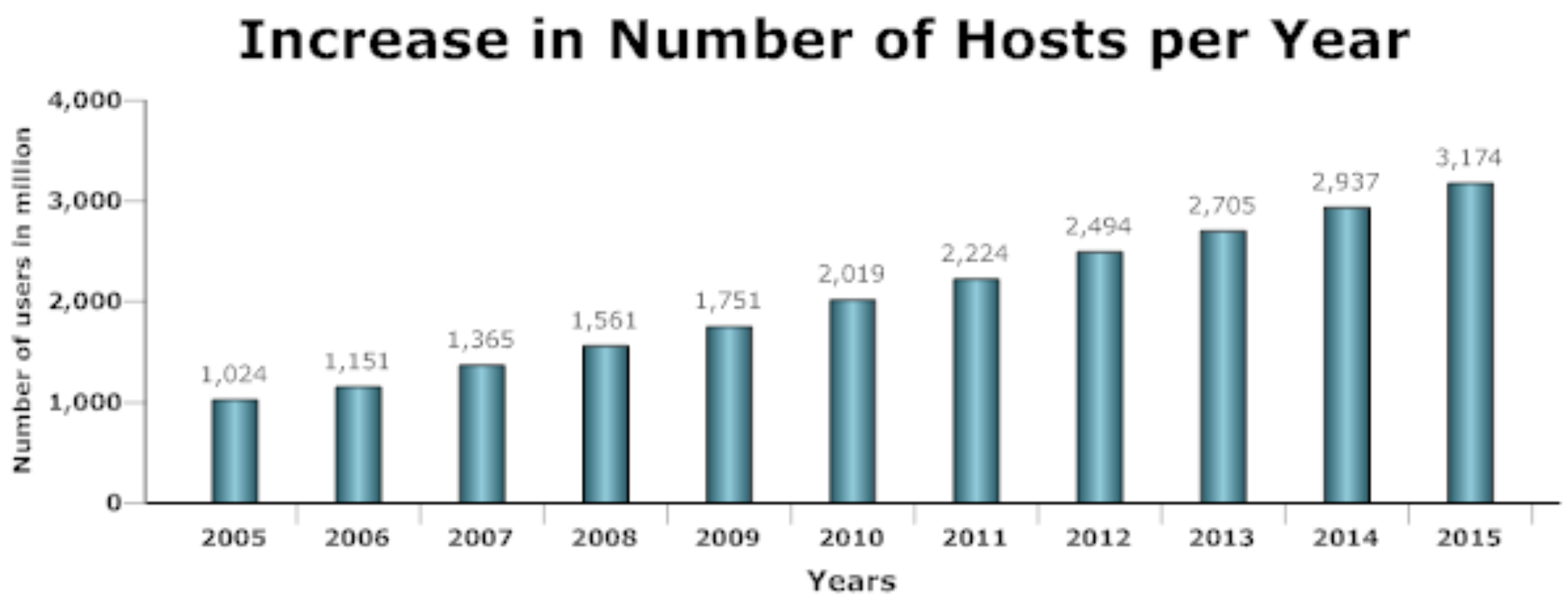

Figure. 1 Growth in number of internet users from 2000 to 2015 [41,42]

Figure 2 shows the size of the Internet users in the world by various geographic regions. This is the recent information according to Internet World Stats [43]. Similarly, figure 3 shows Internet Penetration Rates in the World by Geographic regions according to Internet World Stats [43].

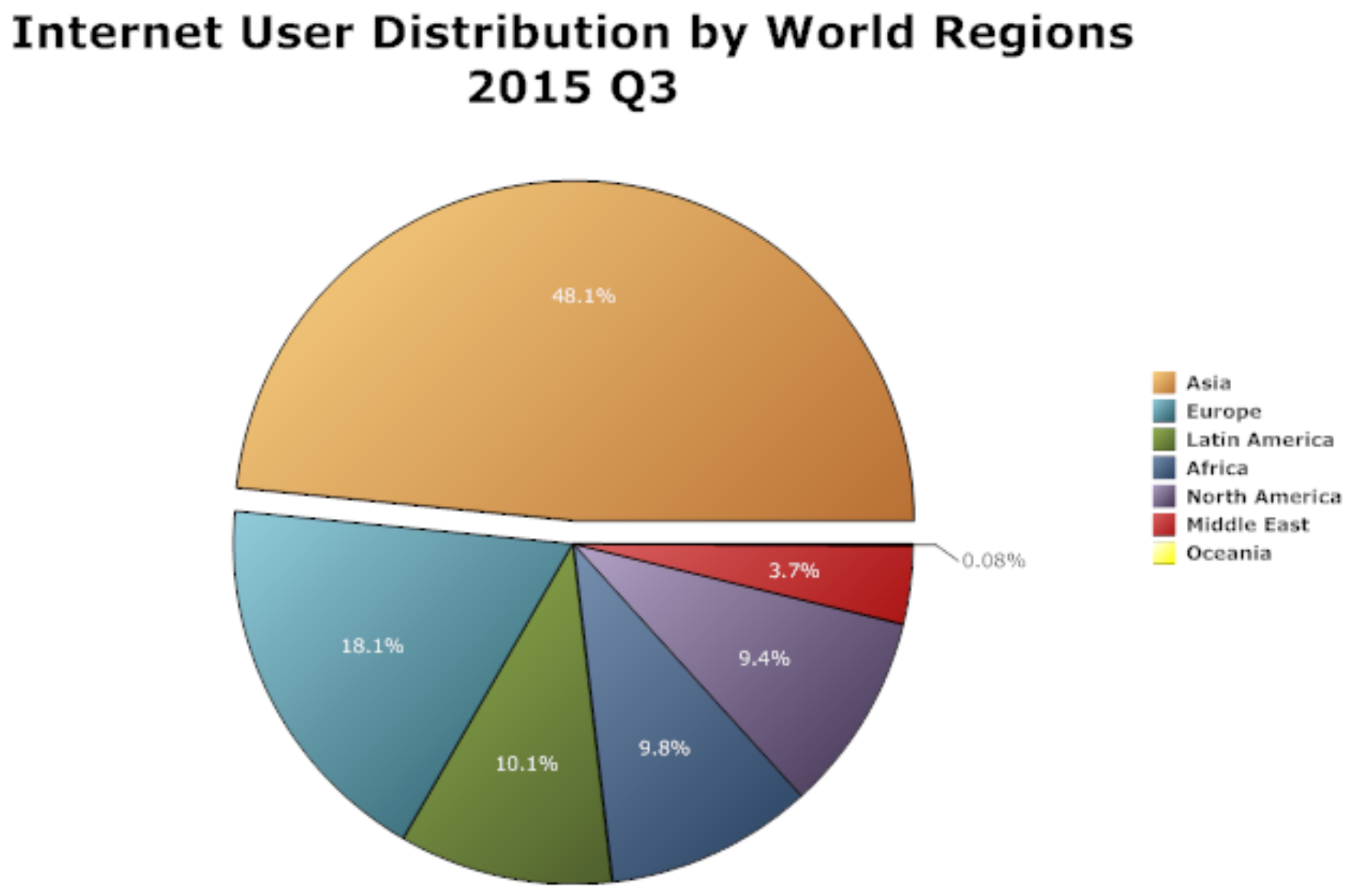

Figure 2. Estimated Internet Users (in Millions) in the World by Geographic regions [43]

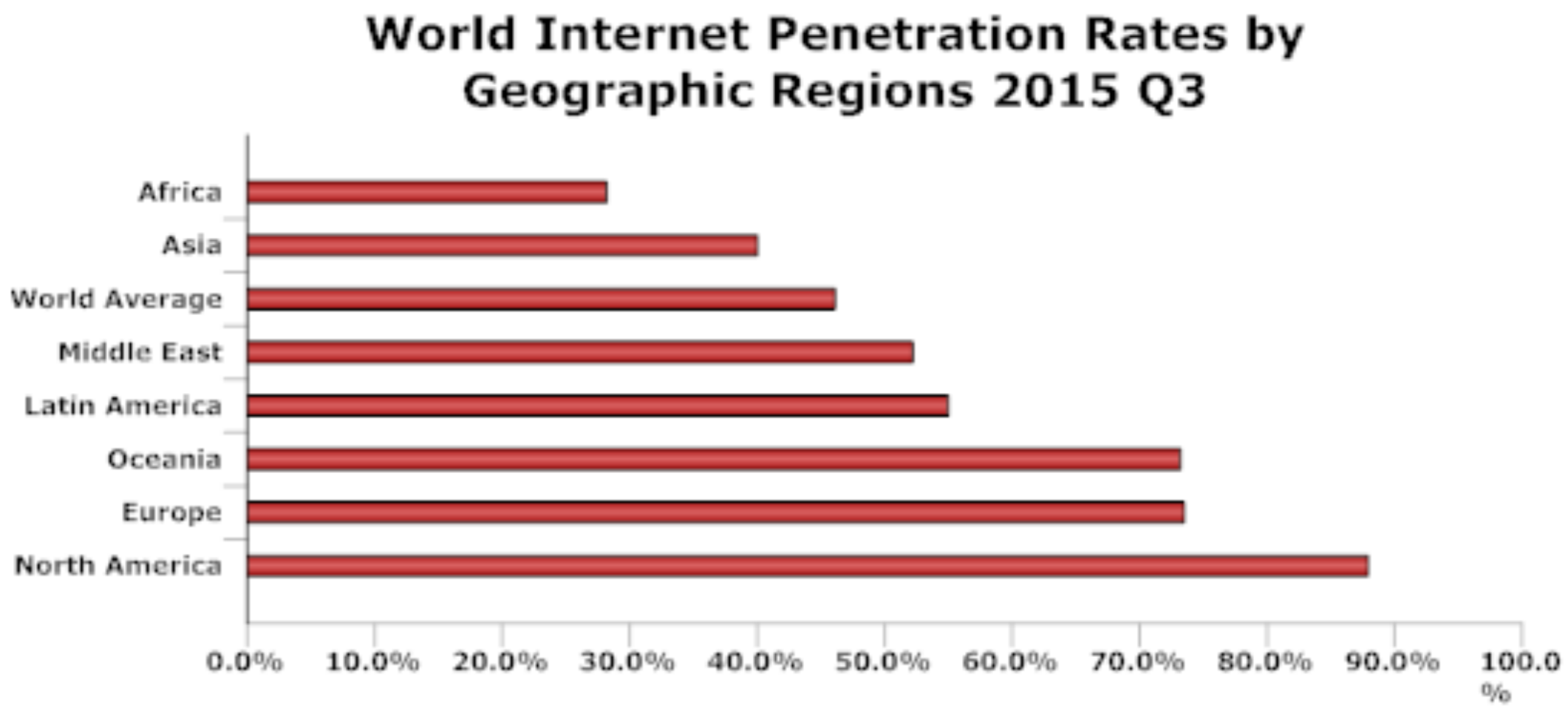

Figure 3. Estimated World Internet Penetration Rates in the World by Geographic regions [43]

Over 250,000 Twitter accounts and over 110,000 job applicant's NPI (National Provider Identifier) were compromised in Virginia Tech's website in early 2013 [66]. In addition, about 74,000 students, staff and faculty members of University of Delaware became a victim of phishing attack and researchers discovered that users' personal details were stolen by an using an existing vulnerability on their website [63]. According to C. Goggi, Phishing attacks were one of the most serious type of threats in 2013 [40]. Malcovery reported that in last quarter of 2013 the top five targeted companies by phishers were Facebook, WhatsApp, UPS, Fargo and Companies House (UK) [44]. Sheng et al. showed that, women were more likely to be a victim of phishing than men. Similar goes for people from 18 to 25 years of age, possibly due to the lack of awareness against phishing threats [45, 56, 61, 62]. According to RSA monthly online fraud reports [65], phishing attack is increasing vigorously over years as shown in Figure 4.

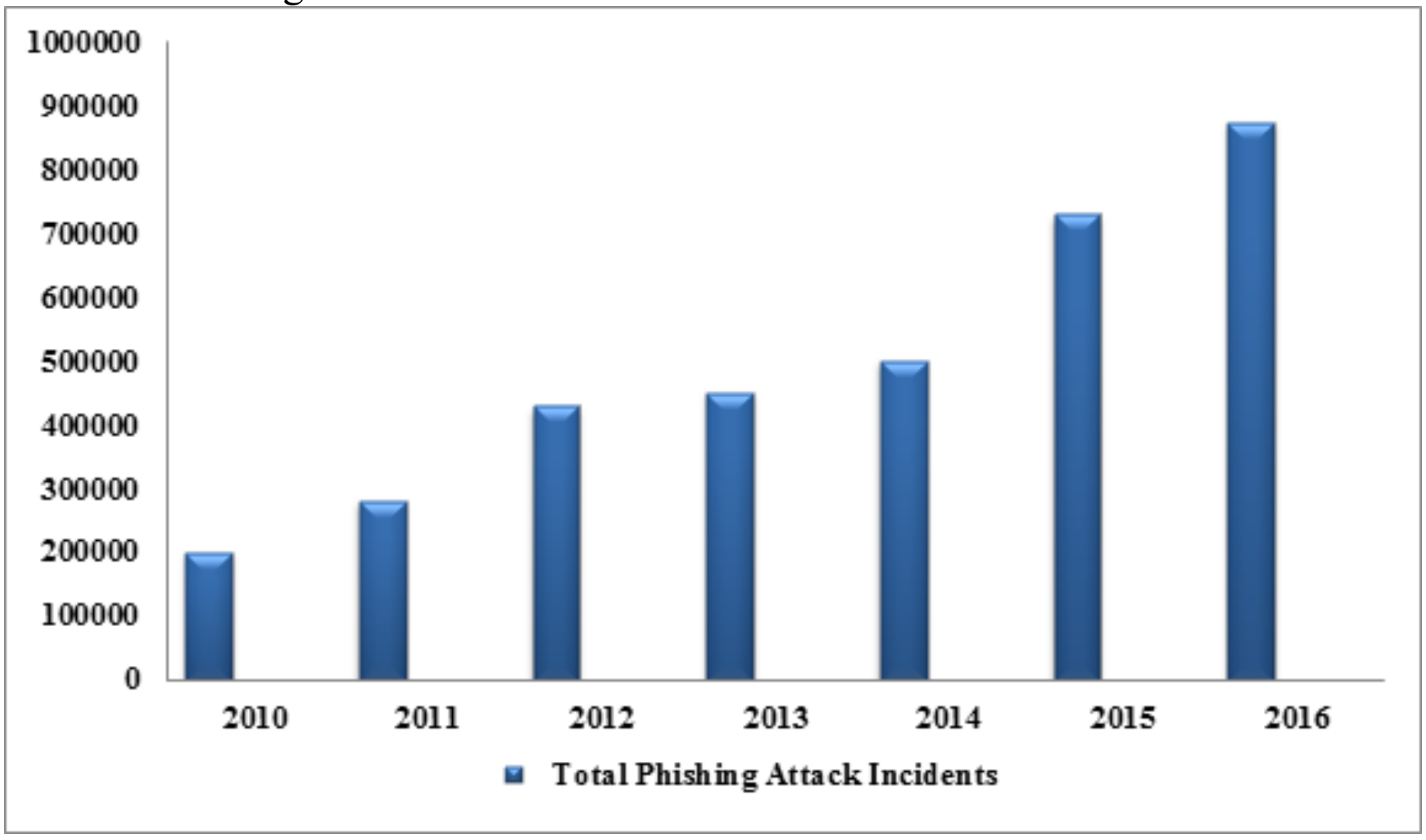

Figure. 4 Phishing attack incidents

The United States Computer Emergency Readiness Team (CERT) gathered security details from various agencies, which stated that there were 107,655 incident in 2011, 43,889 of which were on federal agencies [46]. In May of 2015, construction, engineering, transportation and telecommunication sectors were a target of Advanced Persistent Threat (APT3) phishing campaigns. FireEye identified it to be a zero day attack. The employees received phishing emails having malicious URLs, upon clicking them they redirected to compromised web server, and the target system downloaded an infected Adobe Flash Player SWF file and FLV file which made a backdoor [66].

Hillary Clinton presidential campaign chairman, John Podesta's Google email account was "hacked" in March 2016 prior to the US election [86]. The hacker simply sent a phishing email to Podester's gmail account and lured him to disclose his login credentials. In the phishing email, Podesta had been invited to click on a link (i.e. Unified Resource Locator, so called "URL") warning him to change his password immediately. However, the URL did not link to a secure Google web page, instead directing the user blindly via bit.ly, which is a service used to shorten URLs. Podesta hack didn't require much technical skills. Instead, the hacker merely used social engineering techniques to make the attack successful. The simplicity of the attack, of course, does not make less impact of the crime and makes it no less illegal either.

Therefore, the aim of this paper is to look at the current phishing literature to determine seriousness of the problem. To give a brief overview of evolution of research in this field as well as current trends in phishing and its remedies to provides a view of the issues and challenges that are still prevailing in this area of research. In this paper, we will provide an overview of Phishing problem, history of phishing attacks and motivation of attacker behind performing these attacks. We will also provide taxonomy of various types of phishing attacks. In addition to this, we will provide taxonomy of various solutions proposed in literature to detect and defend from phishing attacks.

Rest of the paper is organized as follows. Section II contains overview of phishing attack (i.e. background, history, phishing lifecycle, motivation, and performance evaluation metrics). Taxonomy of phishing attacks is then discussed in section III. Section IV presents taxonomy of defence solutions. Phishing attacks in the Internet of things (IoTs) are discussed in Section V. Current issues and challenges are discussed in VI. Finally, section VII concludes the paper and discusses the scope for future research work.

## II. Phishing Attack Overview

### *A. Background and History*

Security has been an issue in the field of computer technology since early 50's. In 1950's, the computer had techniques to ensure that a particular application is not able to use memory other than allotted to it. Several encryption and access control techniques to protect passwords etc., were developed in 1960's. Computers were studied as a new complete domain in the 1970's. We have the concept of "Phone Phreaking" since the 1950's till 1980's, that is where the phrase "ph" in "Phishing" comes from replacing "f" in 'fishing' [2]. In 1950, J. Engressia, discovered by accident that certain frequencies can telephone switches with perfect pitch. In 1960, Bell published a paper [47], which included the actual frequencies used for the routing codes. Leak of these codes started a new trend, which was irreversible. In 1964, AT&T began to monitor telephone calls to track phone "phreakers." In 1969, as described in [11] "phone phreaking" was invented by a retired air force technician J. T. Draper. He created a worldwide famous device the 'Blue Box' an electronic device which could use the tones in use by a telephone company so that it was possible to make long distance calls for free, in 1972 he got arrested for toll fraud charges. In 1978, DEC's marketing manager G. Thuerk sent first international commercial spam. A single mass e-mail was written and sent to 393 West Coast ARPANET users for advertising the availability of a new model of DEC computers [2,66]. In 1983, K. Thompson first described a security threat, which is called as "Trojan horse". An electronic magazine named as 'Phrack' which was written by and for hackers, begun publishing in 1985 [2]. A timeline diagram describing various events before Phishing is shown in figure 5.

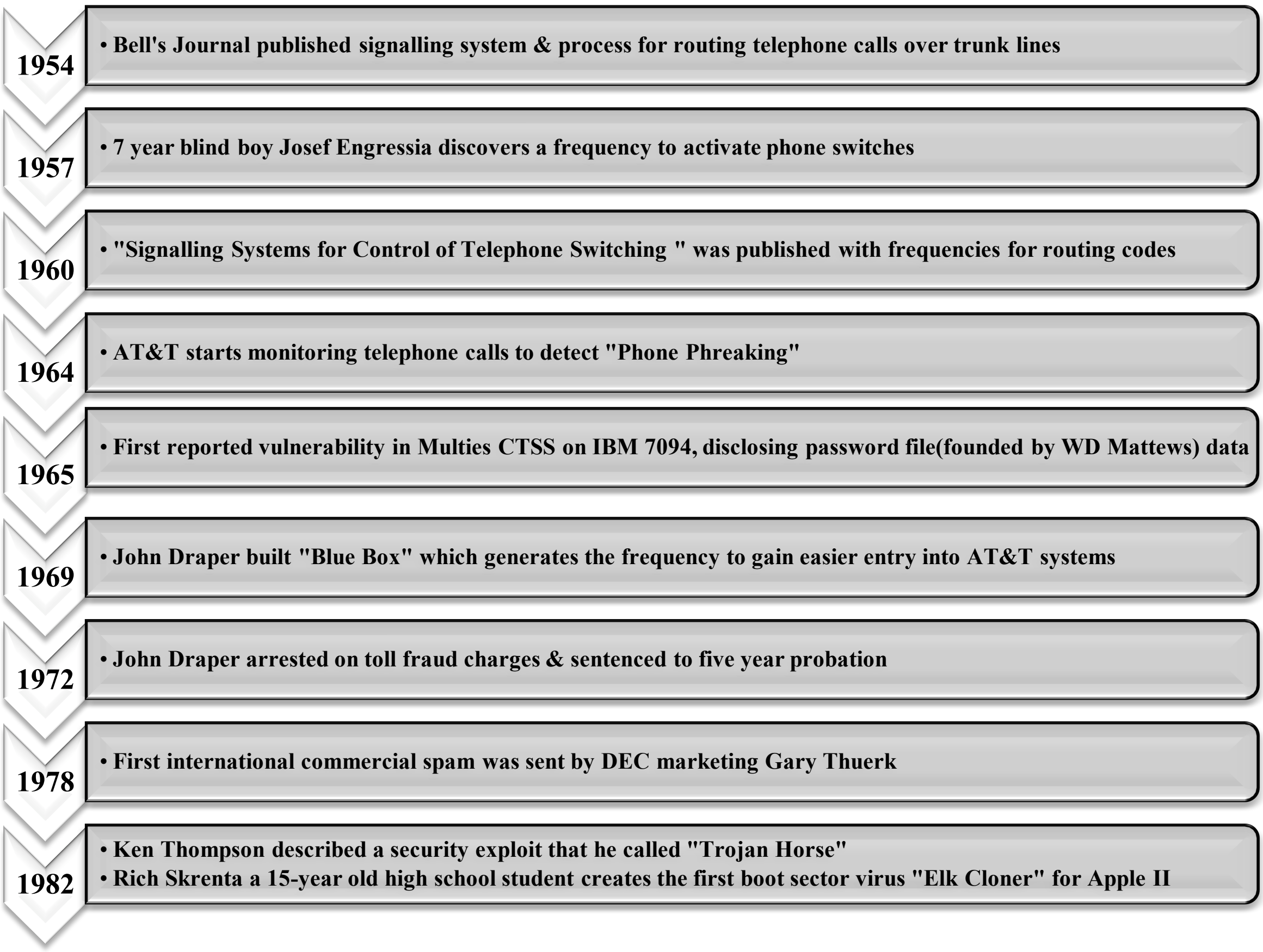

Figure. 5. A timeline diagram describing various events before phishing as described in [5]

We described in detail about the "phishing" era of 1990's and onward in figure 6. In December 1995, it has been reported that hackers attempted to break into DoD (US Department of Defense) computers about 250K times in the same year and 65% of them were successful. In 1996, as described in [11], the term 'phishing' was used first time by hackers who stole America On-line (which is the largest Internet Service Provider in US) by getting access to the passwords of AOL users. As described in [2], phishing was first mentioned on the Internet by the "alt.2600" hacker newsgroup in January 1996, in which hackers asked for any other method to get an account, other than 'phishing'. In addition, in 1997 first media publication warns customers of new threat called "phishing", also AOL cut down its direct access for Russian users due to increased level of fraud. In 1998, phishers started to make use of message boards and news groups to attack victims. From 2000 onwards, phishers started using mass-mailers to spread phishing emails and spoofed URLs to redirect a fake website [2]. In addition to this, for acquiring login credentials (i.e. login-id, password, etc), key loggers became popular among the phishers [2].

In 2001, as described in [11], e-gold became the first victim among the financial institutions. Phishers started using spam messages to spread their network. As described in [11], Buffalo spammer was arrested in New York in 2003 after sending 825 million spam emails and fraudulently using stolen identities. In 2005, Bank of America lost 1.2 million usernames and SSNs of their customers. In 2006, phishers targeted VoIP first time. In 2007, according to Gartner study, about 1.5 millions of US citizen identities got stolen. In 2008, S. Wallace received $711M for posting spam messages on walls of Facebook's members. In 2011, Credit and Debit card details of more than 10M PlayStation Network and Sony Entertainments users are stolen and damaged approximately $1 to $2 billion making it the costliest cyber-hack ever. In February 2014, according to the report of 3rd Microsoft Safer security Index phishing caused annual losses of about $5 billions [48]. Over the past few year, phishing attacks have evolved into much more advanced threats beyond emails also including SMS, online social networking even online gaming [54, 55, 56]. eCrime Trends Reports of the year 2012 shows that Phishing attacks are increasing by 12% per year. Phishing emails are becoming an enormous threat everyday affecting major financial companies and clients. Researchers have given many solutions ranging from authentication protocols to content filtering to protect against phishing attacks but still the attackers are able to carry out these frauds successfully [54, 55, 56, 57]. Of course, it is easy to exploit humans rather than breaking into the system straightway.

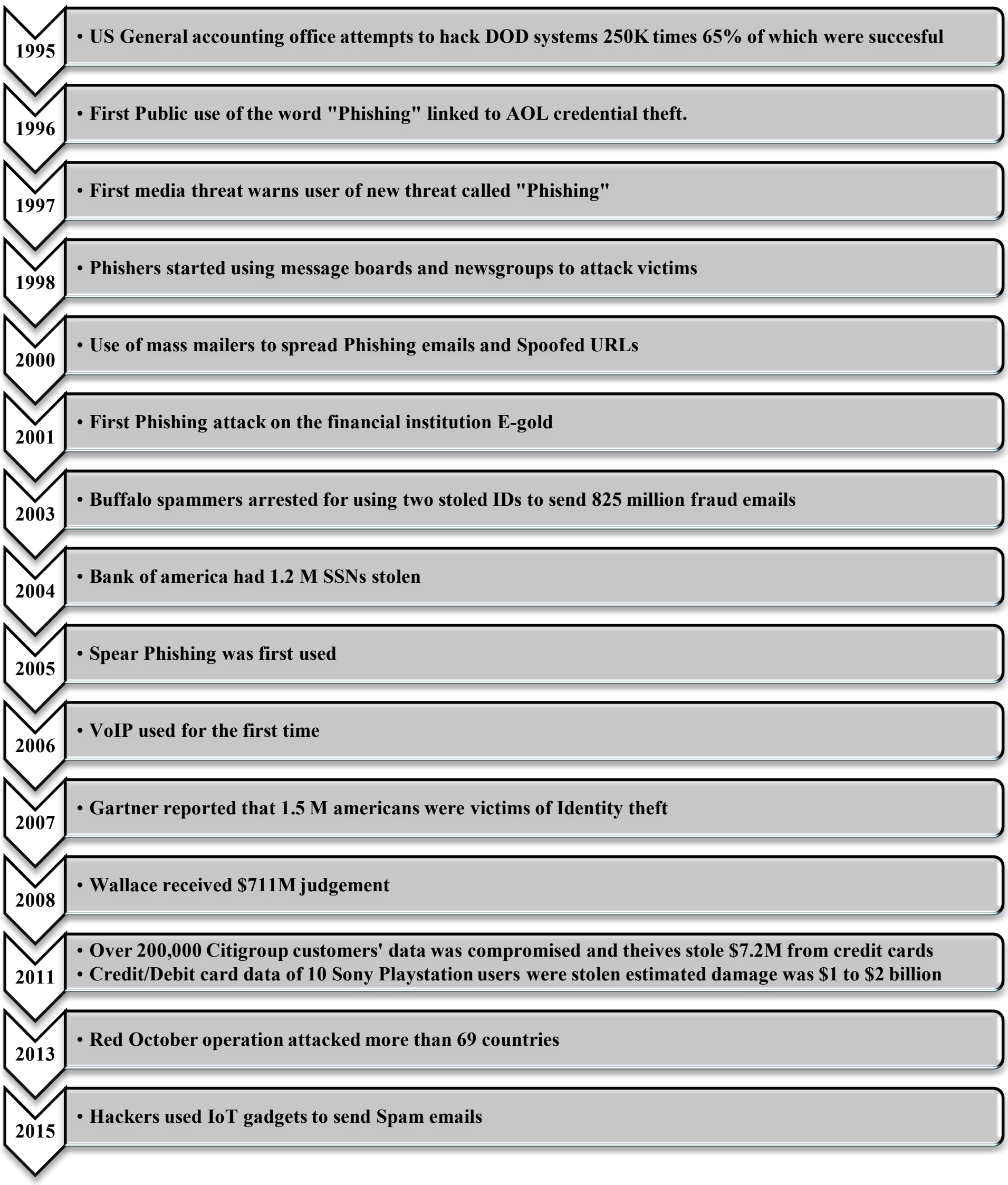

Figure. 6. A timeline diagram of various phishing events as described in [3, 5]

*B. Phishing Statistics Report*

Anti-Phishing Working Group (APWG) is the worldwide coalition of the various governments' law-enforcement sectors. We described the statistics of unique phishing emails and unique phishing websites attacks from fourth quarter 2012 to fourth quarter 2014 in Figure 7 as given in [3]. Based on the statistic reports and figure 8, we can conclude that the number of unique phishing websites attacks in any month is almost double the phishing email attacks, which were unique. Compared to number of unique phishing emails reported, phishing websites attacks which were unique seems to be decreased with time, whereas number of unique phishing emails reported seems to be less decreased in that period.

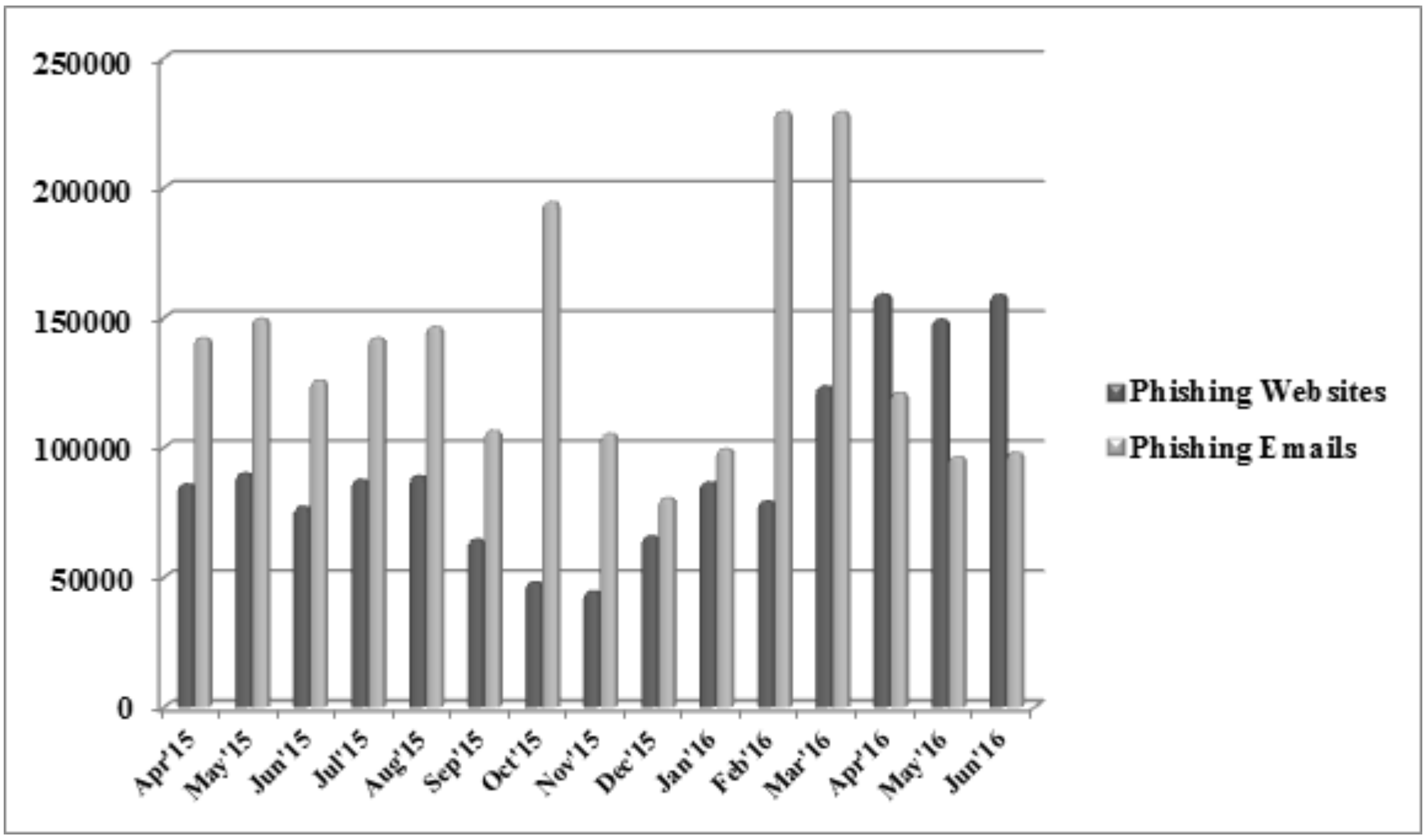

Figure 7. Unique phishing email and websites attacks from January 2012 to December 2014 by APWG

In Figure. 8, we described more about phishing statistics, as given in [3]. We can conclude from these reports that phishers uses port 80 (i.e. http protocol) in maximum cases. In addition, phishers tries to use target name within the phishing URI, so that the page appear legitimate at first glance, even to an expert user. Based on APWG reports, which is available at [3], it is clear that if any solution based on http protocol is able to find the target name from URL and more focused on phishing websites can be able to cover a wider area of phishing network.

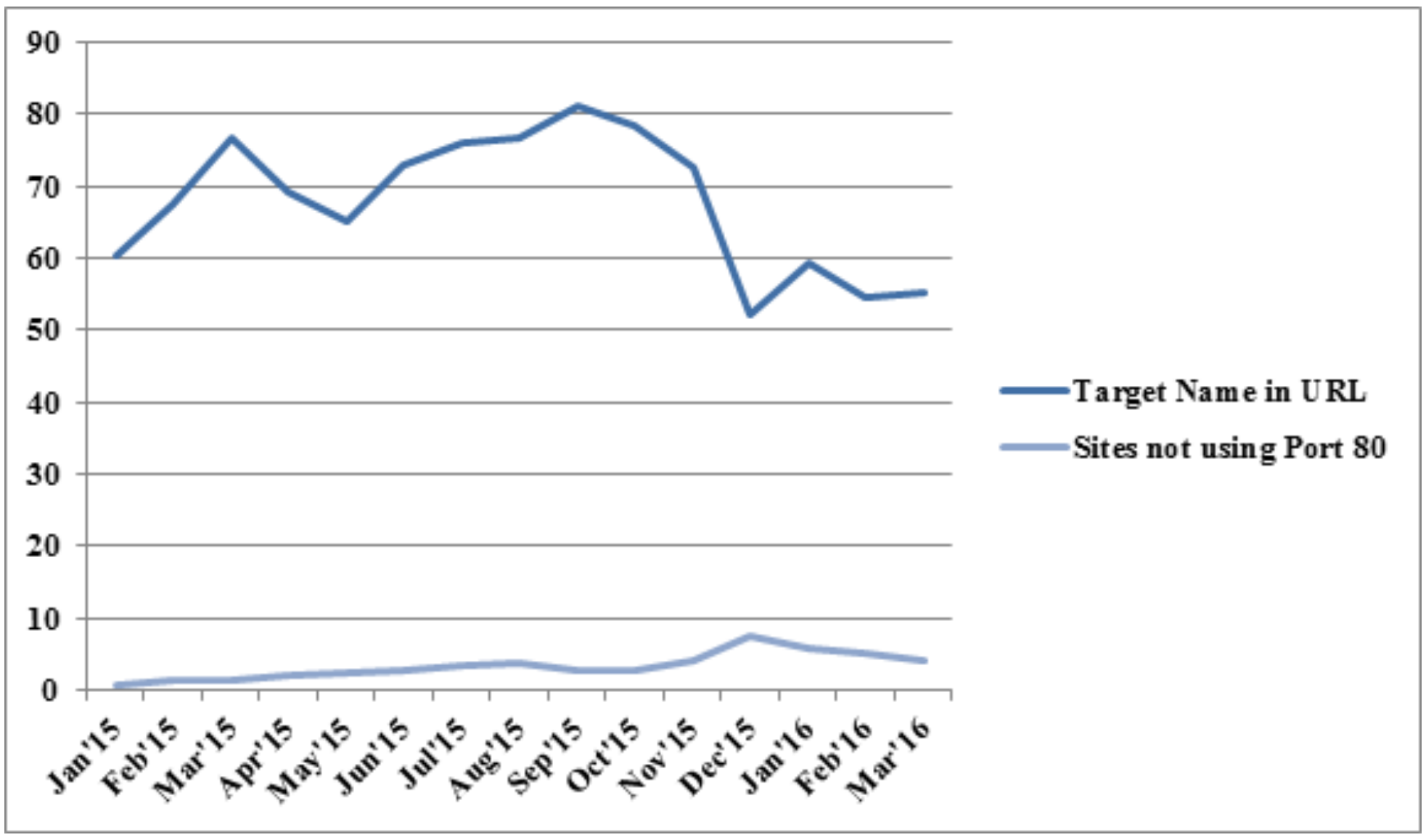

Fig. 8. Phishing Statistics from October 2013 to December 2014

We also studied some statistics of phishing attacks from eCrime Trend Reports. According to [12], in Q4 2014, .com is the most used domain for phishing attacks with 41%, followed by .net with 7%, .org with 5%, .br with 3% and IP address based with 3% (as shown in Figure. 9). Figure 10 shows the details of some highly targeted industry sectors as per the APWG report of 4th quarter of 2014.

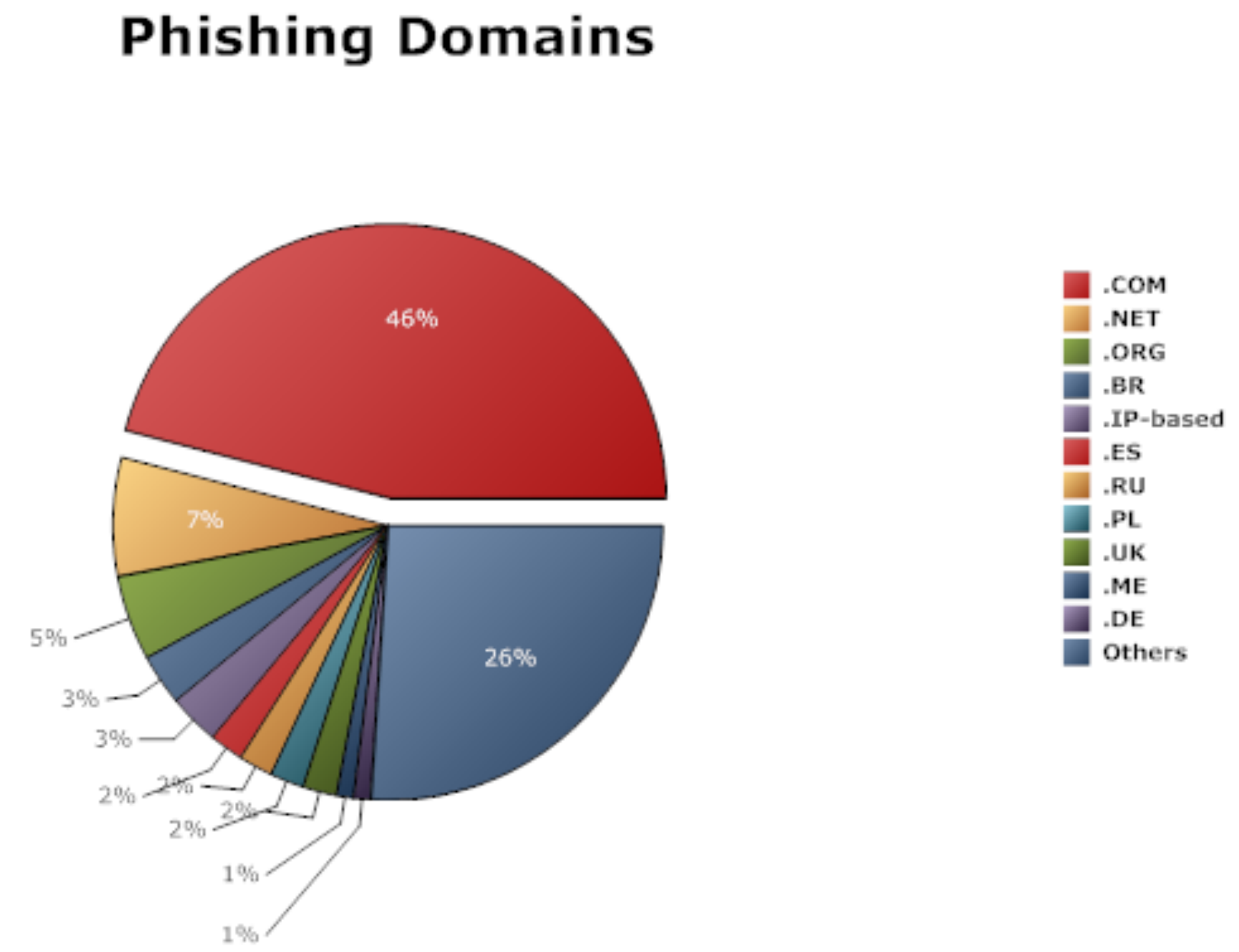

Figure 9: Statistics of phishing websites based on domain (E-crime Report 2013 Q4)

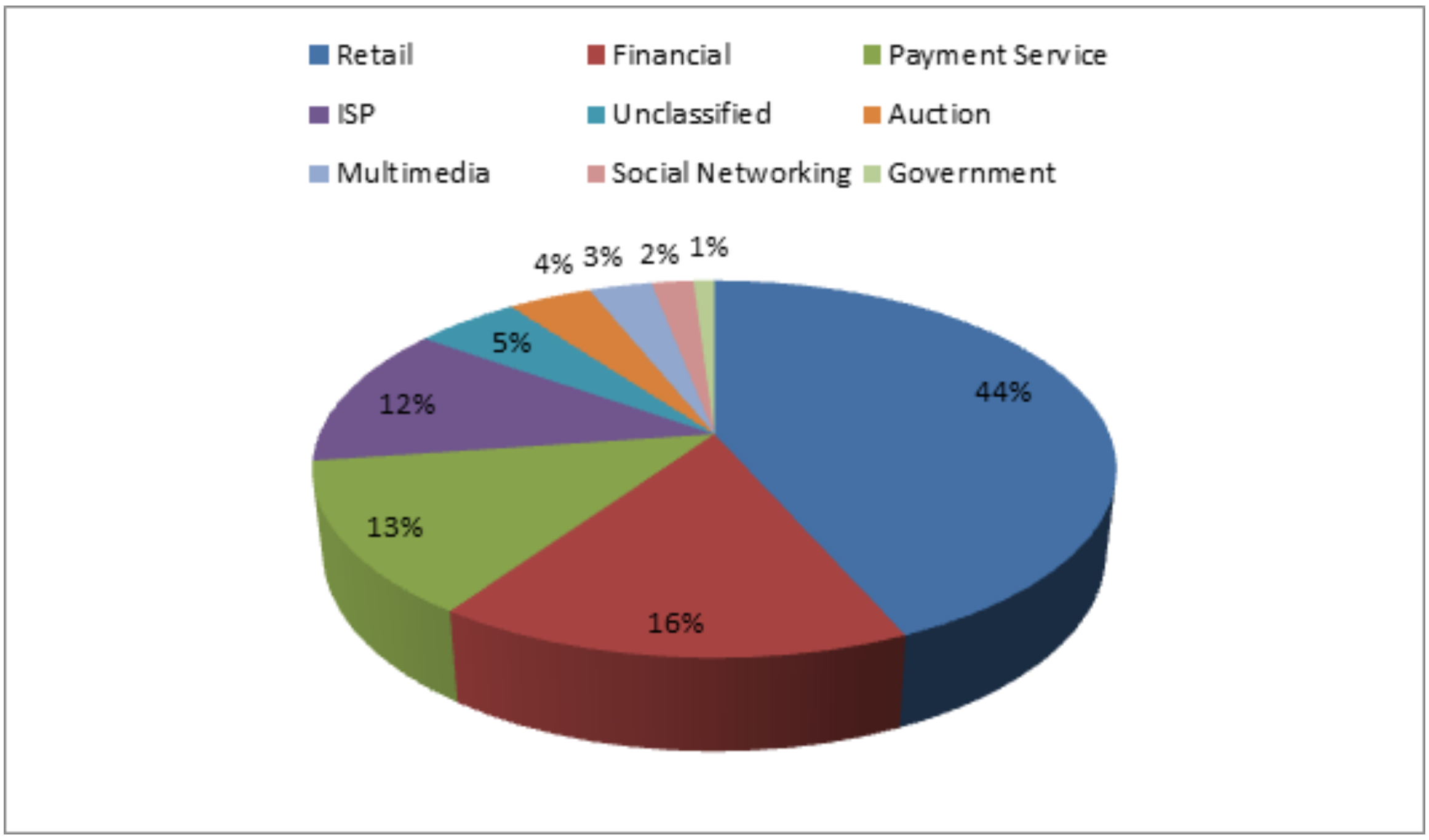

Figure. 10. Statistics showing Most targeted Industries in 2014

*Phishing statistics and incidents of 2015[67-71]*

- Google Safe Browsing mentioned in a report that between years 2014 and 2015, the number of malicious web pages fell down from 18,454 to 14,977 whereas the number of phishing pages rose up from about 24,864 to 33,571.
- Intel security conducted a phishing related quiz online which showed that even trained security specialists sometimes fail to separate phishing and real web pages.
- In a study conducted in 2015, it has been shown that an employee takes about 1 minute and 22 seconds to open a spear phishing email.
- About 70% of the popular IoT gadgets are vulnerable to security attacks.
- Only 6% of the organization claim to have a incident response system and law enforcement.
- In January 2015, three new android families were found, and about 80% of the malware attacks are a part of phishing scams.
- In the year 2014, about 16% of iOS users and 12% of android users fell for phishing attacks.
- One out of every 207 emails contain a virus, and the emails having technical information are most likely to be clicked open.

### C. *Phishing Lifecycle*

There are various phases to the phishing cycle. However, there are three main phases in phishing cycle repeated by various phishers [1, 2, 4-9]. In first phase, the phisher explores organizations and selects a target and then, creates a phishing website and send numerous spam emails among the various users in Internet community. Second phase starts with reading of these emails. Whenever the user "bites" on the phish i.e. click on the link, third phase starts and user is redirected to the phishing site.

In this section, we briefly described about the phishing campaign in which phisher uses the advantage of ignorance about the communication channel in common users, as described in figure 11. In mailing system, every email first passes through the DNS based blacklist filters. If the domain of sender is found in blacklist, the email is blocked before reaching the SMTP mail server. Based on structural properties of emails, various solutions filters email before it reaches to the user's inbox. There are also various solutions available to check emails based on features of any email on client side. In case of phishing webpages, the links are embedded in emails sent to the user or any other advertisement. There are various solutions available on the client side as Internet is vast enough to control it. Some blacklist-based applications block the website if domain falls under blacklist. Unlike the blacklist solution for emails that block emails before they reach the SMTP mail server, it blocks the website when browser of client side request for the URL mentioned in the list. Some more solutions like heuristic feature and visual similarities block the webpage only when the browser request for any phishing webpage [82].

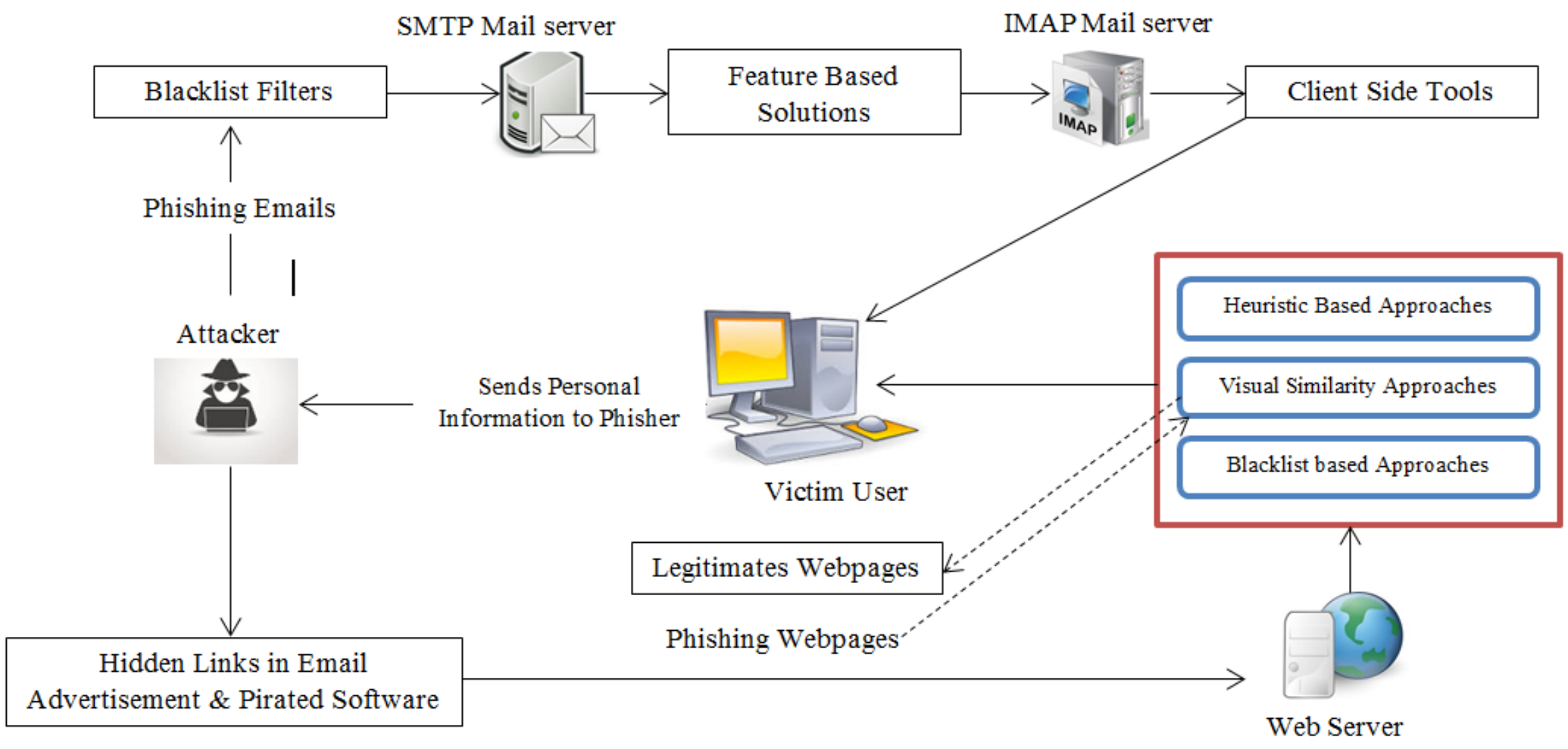

Figure 11. Lifecycle of phishing attacks based on phishing emails and phishing websites

## D. *Motivation*

Phishers always take the benefit of human factors that generally ignore the critical warning messages. Lack of awareness about phishing attacks in society is the main reason, due to which phishing attacks have become so much successful. According to the fact that phishing is mainly used for financial gains, there are other factors such as social gains, motivate phishers to commit the crime. As discussed in [2,7], motivations behind these crimes are as below:

- Theft of banking credentials – stealing of credentials such as credit card details, CVV number and online credentials for websites like PayPal and eBay etc.
- Capture of personal information – personal information, such as address, telephone number are sold online through bids and are in constant demand.
- Theft of trade secrets and confidential documents – here spear phishing is used to accomplish the task, targeting specific organizations for acquisition of proprietary information to use directly or to sell to interested parties.
- Fame and notoriety – sometimes the intention behind phishing scams is not financial gain but to get recognition and fame among peers.
- Exploit security holes – attackers are sometimes curious to discover vulnerabilities in existing or new technologies, which would help them in future to launch attacks.
- Attack Propagation – sometimes to camouflage their location the attackers use bots etc., for propagation, they can use a single system within an organization for running the scam.

## E. *Performance Evaluation*

Performance evaluation is very useful while describing the phishing literature. The efficiency of approaches that are developed for phishing detection are evaluated by one of the metrics given below:

Suppose $N_L$ denotes total number of legitimate pages (or email) and $N_P$ denotes total number of phishing pages (or email). If the legitimate website is correctly identified, it is denoted by$N_L \rightarrow L$. When a phishing website is correctly detected, it is denoted by$N_P \rightarrow P$. Whereas, when the phishing website is detected as legitimate and the legitimate website as phishing, it is represented as $N_P \rightarrow L$ and $N_L \rightarrow P$ respectively. Some of the metrics to measure performance of an approach are mentioned below:

a. True Positive (TP): The ratio of number of phishing pages correctly detected as phishing by the solution with the total number of phishing pages visited.

$$TP = \frac{N_P \rightarrow P}{N_P}$$

b. True Negative (TN): The ratio of number of legitimate pages correctly detected as legitimate by the solution with the total number of legitimate pages visited.

$$TN = \frac{N_L \rightarrow L}{N_L}$$

c. False Positive (FP): The ratio of number of phishing pages wrongly detected as legitimate by the solution with the total number of phishing pages visited.

$$FP = \frac{N_P \rightarrow L}{N_P}$$

d. False Negative (FN): The ratio of number of legitimate pages wrongly detected as phishing by the solution with the total number of legitimate pages visited.

$$FN = \frac{N_L \rightarrow P}{N_L}$$

For simplicity we can considered that 'True' represent the correct detection, whereas 'False' represent the wrong detection. 'Positive' represent that the actual page is phishing, whereas 'Negative' represent that the actual page is legitimate. However, these are not the only measures to evaluate performance. Some more evaluation measures given in [7, 8] are described below:

e. Precision (P): The ratio of number of phishing pages correctly detected as phishing by the solution with the total number of phishing pages detected by the solution. The phishing pages detected by the solution may be actually a phishing page or a legitimate page.

$$P = \frac{N_P \rightarrow P}{N_P \rightarrow P + N_P \rightarrow L}$$

f. Recall (R): The ratio of number of phishing pages correctly detected as phishing by the solution with the total number of actual phishing pages visited by the solution. It is simply the True Positive of any solution.
$R = TP$

g. $f_1$ Score: It is the harmonic mean between Precision and Recall. It tends to take into account the performance of the algorithm on common categories.

$$f_1 = \frac{2\mathrm{PR}}{\mathrm{P} + \mathrm{R}}$$

h. Accuracy (ACC): The ratio of sum of correctly detected legitimate and phishing pages by the solution with the sum of total number of actual legitimate and phishing pages visited by the solution.

$$ACC = \frac{N_P \rightarrow P + N_L \rightarrow L}{N_P + N_L}$$

i. Weighted Error (WER): The ratio of sum of incorrectly detected λ times the number of legitimate pages and phishing pages by the solution with the actual number of phishing and legitimate pages visited by the solution.

$$WER = 1 - \frac{\lambda. N_L \rightarrow L + N_P \rightarrow P}{\lambda N_L + N_P}$$

Another important category is for the evaluation of features used in email classification. The most commonly used metrics for this purpose are:

a. Entropy (E): Measures the amount of disorder or disturbance in the system. It can be calculated as:

$$E(S) = \sum_{i=1}^{N} -p_i log_2 \, p_i \ , \qquad (10)$$

where, N: number of classes in the dataset,
S: dataset, and
$p_i$: probability of an email belonging to class i.

b. Information Gain (IG): Measures decrease in the value of entropy when a particular feature is used. IG(S,A) is the information gain of dataset S over the attribute A and can be obtained as :

$$IG(S, A) = E(S) \ - \sum_{v \in value(A)} \frac{S_v}{s} E(S_v) \ , \qquad (11)$$

where, $S_v$: the number of attributes in S with A has the value of v, and.
$E(S_v)$: entropy of the subset $S_v$ in S.

## III. TAXONOMY OF PHISHING ATTACKS

Phishing attacks can be classified based on the mechanism by which the phisher can be able to access personal information of a victim. Either a phisher uses a way in which he/she frauds to victim or he/she uses any malicious code to access victim personal information. A phisher may fraud to any innocent user either by using spoofed emails or by using fake websites. Malicious code, key logger, and screen capture can also use to access personal information and technique is known as technical subterfuge phishing. A basic classification of Phishing attacks is shown in Figure. 12.

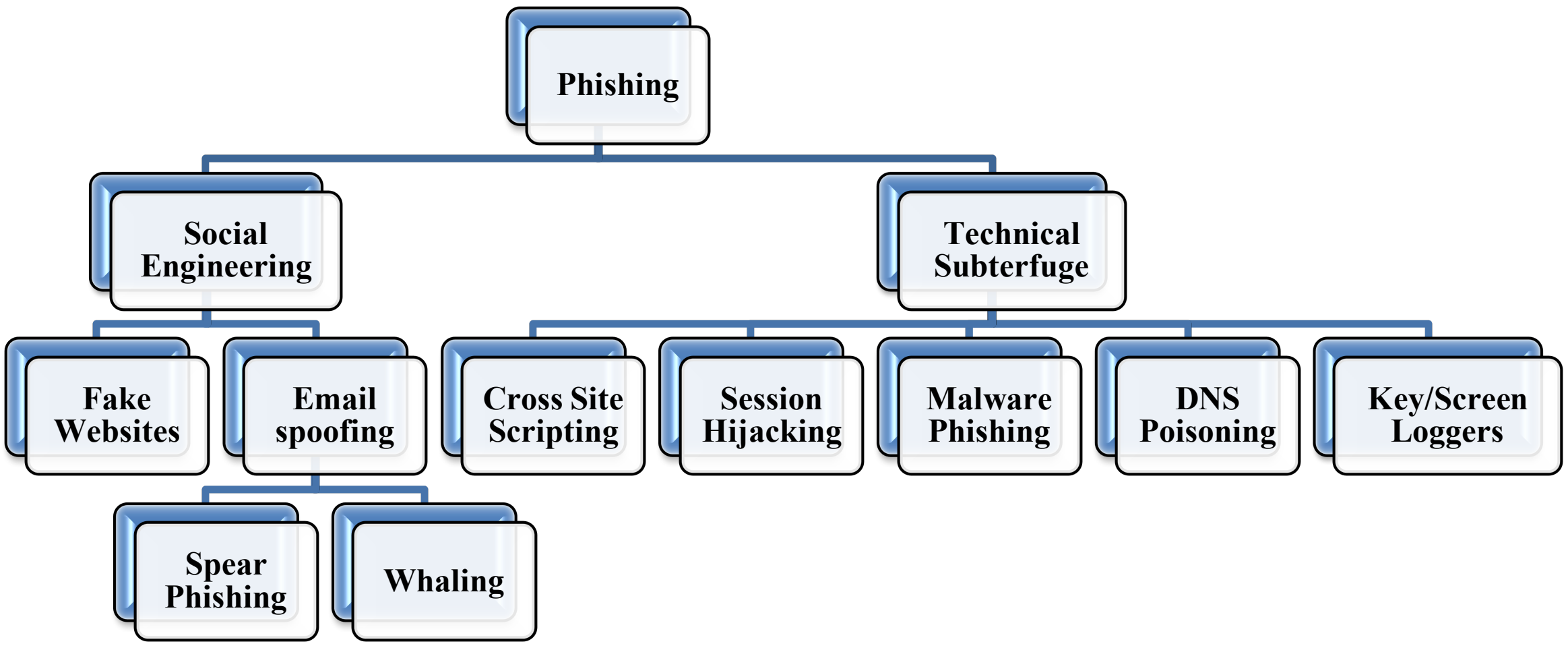

Figure 12. Classification of Phishing attacks based on how a phisher fraud to victims

### *A. Social Engineering*

Social engineering refers to fooling a person to accomplish certain goals, which may be malicious and harm the victim. These techniques basically intend to acquire access to certain system or getting information related to a person or a group for financial gain. During 2014, Apple was the most targeted brand by the phisher according to global phishing survey. Attackers sent fake emails to apple users which asked them to update their account details, giving a link to redirect them to website where they could update the data, several users ended up giving their credential on those fake sites [72]. As defined in [3]:

"Social engineering schemes use spoofed e-mails purporting to be from legitimate businesses and agencies, designed to lead consumers to counterfeit websites that trick recipients into divulging financial data such as usernames and passwords." The phishing based on social engineering is further classified based on [14-19]:

I. *Spoofed Emails*

a) We also called them phishing emails. These emails are not as ordinary emails, but these are from untrusted mail server or some pranks are used to make believe its victim that any trusted party sent this mail. These emails are used to convince its victim, so that he/she may send his/her personal information. Email phishing can be done in any of the following ways: (i) concatenating some string to start or end of a legitimate domain to generate a fake link; (ii) the actual links are not the same as the links visible to the user; (iii) use bugs to redirect a link to a malicious website; (iv) exchange certain characters of legitimate URL with similar characters that are different to detect; (v) Use of Javascripts etc, to hide the address bars.Spear Phishing- A new term "spear phishing" has also come into picture, where the target is a specific person or organization. Spear-phishing is also being used against any group of people in an organization working at any position, "whaling" specifically targets the high-rank employees of an organization [64]. A spear phishing attack targeting to a specific user may leverage information such as his/her user name and email address to craft an email that is personalized to the user. This spear phishing technique will certainly improve the success rate of the attack and techniques that can be leveraged by an attacker to find contextual information [87]. In the year 2009, major organizations such as Google, Yahoo, Adobe and Symantec became victims of spear phishing and malware attacks by a group Operation Aurora attacks.

b) Whaling- This mainly targets high profile employees of big organizations to excess highly confidential information. It is also called CEO fraud, here hackers use social engineering to phish users to give away their bank credentials employee data etc. These attacks are even difficult to detect as they do not use malware or fake websites.

II. *Fake Websites*

These are also known as phishing websites, they appear same as that of a legitimate site visually, these fake websites are used to get personal information about the victim. Generally, the links of the fake websites are embedded with phishing emails, advertisements or within crack of licenced software. Phisher always try his best to make visual appearance and URI pattern similar to the victim site.

B. *Technical Subterfuge*

A phisher not just depend only on the social engineering schemes to theft personal information. Technical subterfuge is another popular way to fraud in which a phisher send some malicious code either attached with emails, or with websites, or with some self-executable code (generally crack of any software). As defined in [3]:

"Technical subterfuge schemes plant crime-ware onto PCs to steal credentials directly, often using systems to intercept consumers online account user names and passwords and to corrupt local navigational infrastructures to misdirect consumers to counterfeit websites." The phishing based on technical subterfuge is further classified based on [14-19]:

a) Cross-Site Scripting – XSS is a vulnerability due to weak security techniques in web applications. To bypass the access control, attacker may be used cross-site scripting vulnerability. XSS happened when dynamic web page displayed input without properly validate. This allows attacker to install malicious JavaScript into generated page that is viewed on victim side. After this, it is very easy to transfer login credential to the attacker and misuse for financial gain or any other purpose. XSS can influence any site that allows the user to enter data.
b) Session Hijacking – Session hijacking is a common and serious thread in WLAN. This is also known as cookies hijacking. In this attack, session key is hijacked with the help of denial of service attack to steal the identity and access the unauthorised resources. An attacker force mobile station to terminate its connection with any particular access point. It have been done by disassociate the source MAC address of current access point and spoofed the access point to attacker.
c) Malware Phishing – In malware based phishing, malware is used to store credentials in victim computer and send it to the owner i.e. the phisher.
d) DNS Poisoning – In DNS poisoning attacks the phisher has a fake DNS server and somehow tempts the client to communicate with it, and once the victim connects they are directed to malicious webpages or might install malware into their systems.
e) Key/Screen Loggers- Key loggers are very difficult to detect and now with screen logging software the virtual keyboards have no utility at all. These capture the screen snapshots and mouse movements which are sent to the phisher who is at a remote location.

# IV. TAXONOMY OF DEFENSE MECHANISHMS

We classified various phishing detection and protection schemes based on some classifications describe in [14-19]. One of the classifications is used to classify various available solutions based on email filtering schemes [19] and other classification is used to classify various available solutions based on detecting phishing websites [16-18]. In email filtering classification, there are some schemes like network based protection that is based on blacklist schemes, heuristic schemes in which phishing emails either detect on server side or on client side, are based on some features which is introduced by phisher to redirect victim to phishing websites or other features used to fool victim. In phishing websites detections, some extensions or toolbars are used with web-browser to protect from phishing attack. There are both server side (CANTINA, PILFER etc.) and client side solutions (balcklisting and whitelisting) available in email based filtering method, but during our literature survey, we only found client side solutions in websites detection schemes.

## *A. User Education*

*Why do people fall for phishing?:* Dhamija, et al. [91] conducted a laboratory-based study showing twenty-two participants to twenty websites, asking them to differentiate phishing website from legitimate ones. Authors revealed that that participants made mistakes on the test 40% of the time. Furthermore, authors noted that 23% of their participants missed out all cues in the web browser address bar and status bar as well as all security indicators. Nevertheless, to date, a considerable amount of literature work has been discussed that "humans' incapability to interact with the systems" is one of major reasons why people still fall for phishing attacks [92, 59, 60, 61, 62, 93]

It is vital to state that users' perception of such phishing threats may encourage users to prevent from potential vulnerabilities. Downs, et al [92] have employed a role-playing study aiming at understanding why people fall for phishing emails and what cues they look for to prevent such attacks. The results revealed two key things. First, even though people are aware of phishing attacks, they do not feel potential vulnerabilities or strategies to trace phishing attacks. Second, while people can protect themselves from known risks, they also have difficulties of understanding their known to unfamiliar risks. Wu, et al. [93] stressed that most users do not understand how phishing works or how sophisticated such attacks can be. One could argue this would be the people's lack of phishing threat perception.
Wu, et al [93] empirically investigated three simulated anti-phishing toolbars to determine how they were effective at securing participants from visiting fraudulent websites. Their study revealed that many participants ignored passive toolbar security indicators and instead used the website's content to decide whether or not it is a phishing website. In some cases, participants did not notice warning signals and in other cases they assumed warnings were not valid though they noticed them. Perhaps, one can argue people struggle to interact with toolbars due to a lack of usability. Therefore, it is worth understanding to conduct usability studies to emphasis how users interact with security toolbars.
Previous research has been shown that both academic and government organizations have made a significant effort to deliver end-user education to enable public understanding of the importance of cyber security, especially in anti-phishing context [94]. The Anti-Phishing Work Group (APWG) [3] is a non-profit organisation working to provide anti-phishing education to improve the public understanding of computer security. They cover number of areas: 1) What a phishing threat is ? 2) How can it be severe? 3) What is the usefulness of having a safeguarding measure? 4) Where and how to report a suspected phishing email or website? and 5) Anti-Phishing education to thwart phishing attacks. The US Computer Emergency Readiness Team also offers people free advice on its website about common cyber security breaches for computer users who have a limited computer literacy.

While a great deal of efforts have been contributed to resolve the phishing issue by prevention and detection of phishing techniques related to emails, URLs and web sites, little research has been done in the area of end-user education to protect themselves from phishing threats [94]. Therefore, further research needs to aim at anti-phishing education to protect people from phishing attacks.

### *B. Protection from Phishing Emails*

Phishing email message transportation is shown in Figure 13. The framework to detect phishing emails from a set of emails in real time situations is present between Message Transfer Agent (MTA) and Mail User Agent (MUA) to stop phishing email before reaching the victim's inbox.

*MTA (Message Transfer Agent):* acts as mail carrier and storage location.
*MUA (Mail User Agent):* application that retrieves emails such as "MS Outlook".
*MDA (message delivery agents):* it is the mailbox, it saves messages until the user sees them.
*Phisher:* person or group with malicious intentions targeting a victim.
*Victim:* user who may fall for the phishing email and become the target of Phisher.

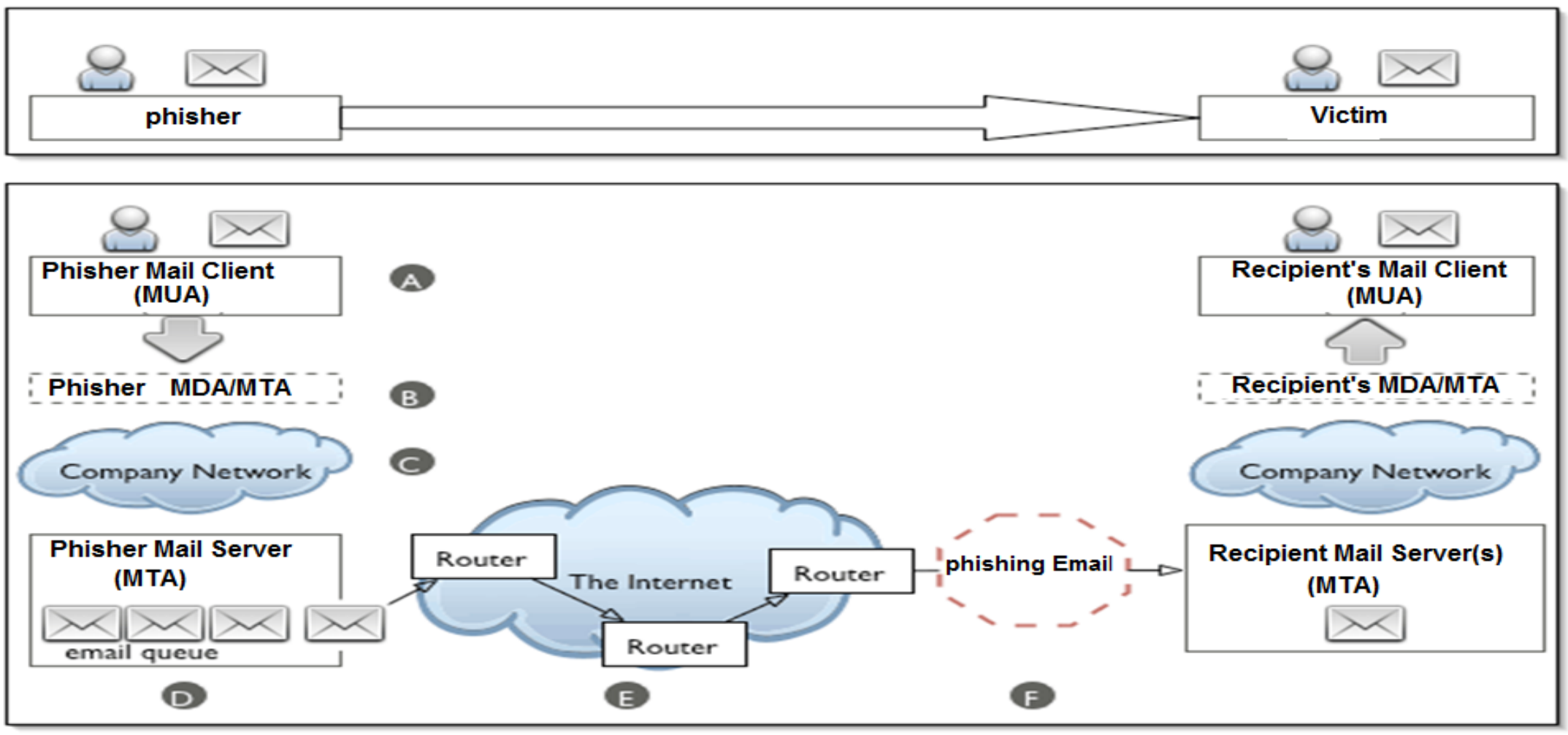

Figure 13 Phishing email message transportation

An overview of email data parts is shown in figure 14 [49].

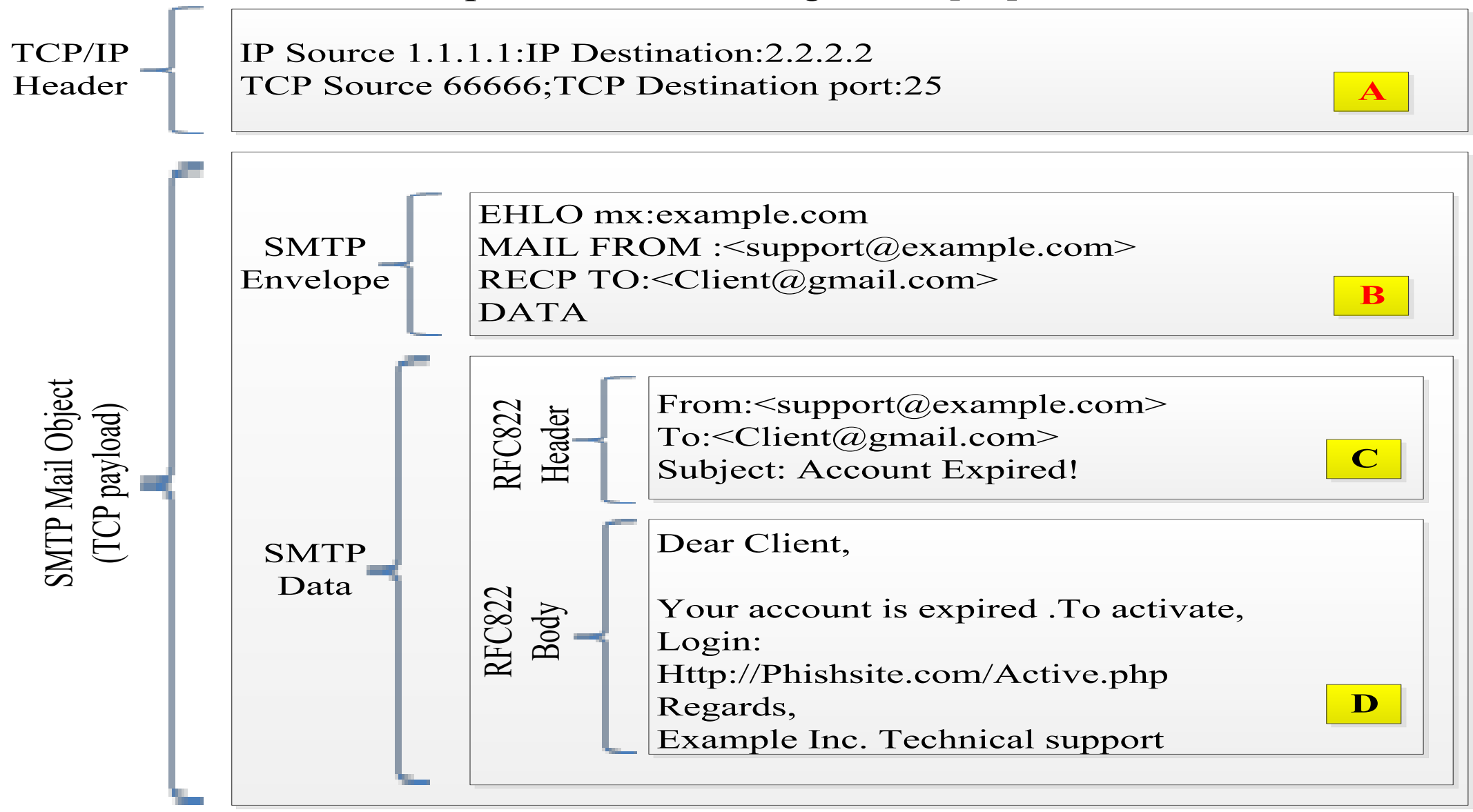

Figure 14: An overview of email data parts [49]

Phishing email Features:

The most effective features are extracted by analyzing data parts C, and D of an email message (as shown in Figure 14). A generally used approach in extracting features found in A and B is by the use of blacklists [49]. The groups of most effective features of an email are discussed in Table I.

Table I. Most effective features of an Email

| **Group Features** | **No** | **Features** | **abbreviation of Features** |
|---|---|---|---|
| **External features** | 1 | Spam features (included 50 sub-features) | Spamfeatures |
| **Body-based features** | 2 | HTML e-mail | body_html |
| | 3 | Body of Multi part | body_multipart |
| | 4 | Verify your account phrase | body_Verifyphrase |
| | 5 | “OnClick” JavaScript event | body_JSonclick |
| | 6 | Code of JavaScript to Change the status bar | body_JSchangebar |
| | 7 | Code of Java script | body_javascript |
| | 8 | Code of Java script to open popup windows | body_JSpopup |
| | 9 | Forms in email body | body_forms |
| | 10 | Ratio of the number of words to the number of Characters | body richness |
| **URL-based Features** | 11 | html-links | url_htmllink |
| | 12 | Number of dots in a link | url_nodots |
| | 13 | Non matching between target and text of urls | url_ TarDiflink |
| | 14 | URL IP address | url_ip |
| | 15 | Image links | url_imagelink |
| | 16 | URL bag of word links | url_bagword |
| | 17 | URL has two domains | url_twodomain |
| | 18 | Non-standard port in the URL | url_nonstport |
| | 19 | URL containing hexadecimal characters or @ symbol | url_hexorat |
| **Header based Features** | 20 | Subject replay word | sub_replay |
| | 21 | Difference between the sender domain from the domain of the embedded links | Diffsenlindom |
| | 22 | Subject (bank, verify, debit) | sub_words |
| | 23 | Sender e-mail address uses different replay address | Sendiffreplyto |
| | 24 | Total number of words in the subject line | subj noWords |
| | 25 | Total number of characters in the email's subject | subj noCharacters |
| **Sender based Features** | 26 | Total number of words in the send field | send noWords |
| | 27 | Total number of characters | send noCharacters |
| | 28 | difference between the sender's domain and the reply-to domain | send diffSenderReplyTo |
| | 29 | sender's domain is different from the email's modal domain | send nonModalSenderDomain |

Table I shows four groups of features: External Features (group 1), Body-based features (group 2), URL based Features (group 3) and Features Header (group 4). Phishing emails are the traditional and one of the common ways for phishing frauds. Users through Mail User Agent (MUA) transfer any phishing mail from Mail Transfer Agent (MTA), which transferred email to Mail Delivery Agents (MDA), and then finally received. The Figure. 15 shows the process of phishing email being transferred to a computer network [19].

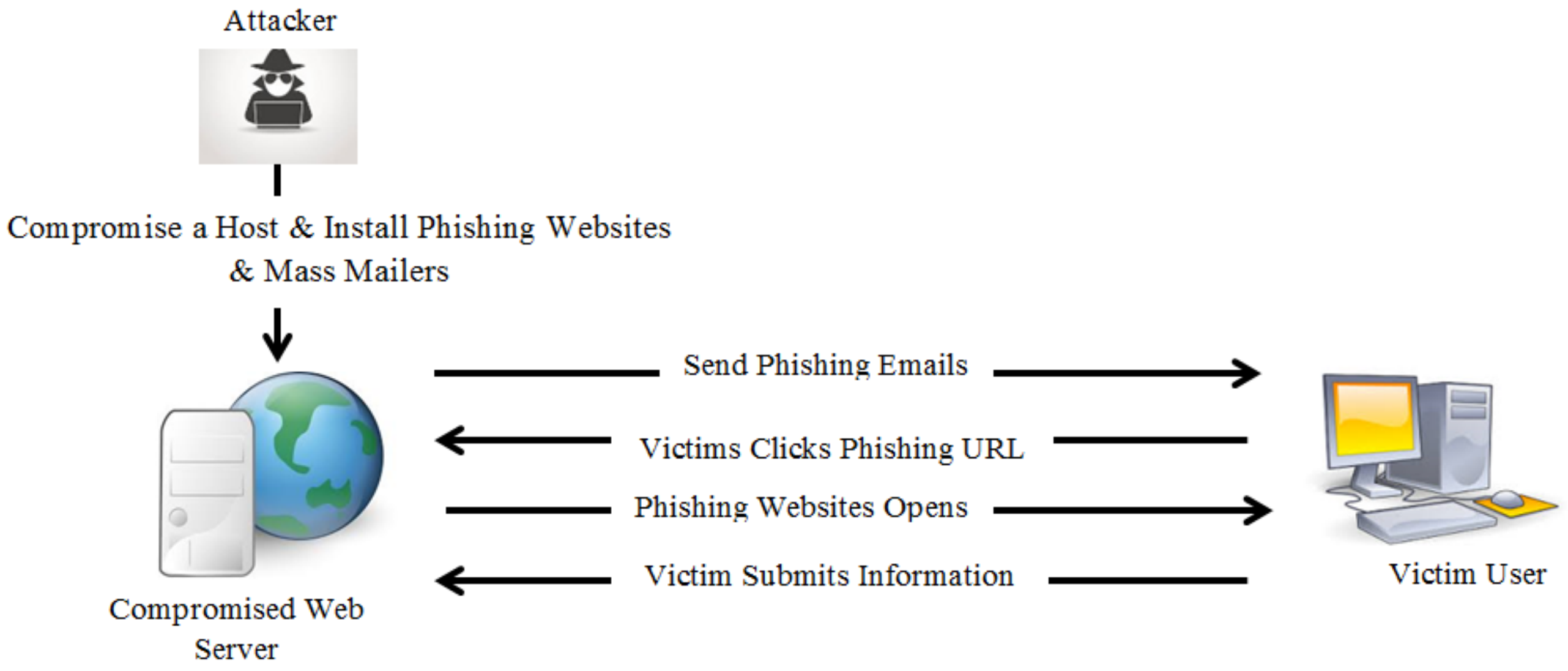

Figure 15. Various phases in phishing email attack

Based on life cycle of phishing email, as discussed in [10] phishing emails are classifying into the following categories.

a. *Network level protection*

Implementation of network level protection restricts a set domains or IP addresses to enter in a network. It blocks the communication from those systems, which are marked or identified as spammers or phishers. The network level protection is also known as 'blacklist filters', because it based on the mechanism in which some particular range of IP addresses or domain name listed as blacklist and not allowing any communication from the list. Some example of this filter schemes as given below:

i. Anti-Spam filters

The anti-spam techniques [19] can determine the origin of emails and decrease the attacks to great extent. Emails are sent in bulk to mark potential victims; these emails contain fake sender details and a false route information etc. Due to the vulnerability that exist in SMTP protocol all of this is possible. Emails to seem to be coming from a genuine organization and phishers are able disguise their actual identities.

ii. DNS-Based Blacklist

The blacklist approaches are reactive in nature requiring attention from Internet Service Providers to continuously update the list by monitoring the network traffic. The DNS-based blacklists [20] make use of DNS protocol. However, a server optimized to handle large DNS resource records is required or else other service handling DNSBLs may face several limitations in terms of performance and speed. However, this technique has also been cracked by the attackers by acquiring access to a legitimate system or using different IP addresses.

*b. Authentication*

User and server authentication approaches check whether the attacker is not pretending to be a valid sender of an email or a resource request, it increases the security at both server and user level. At user level authentication is ensured by use of passwords, but it is evident in past that passwords can be cracked by the phishers [21]. Authentication at domain level [22] is ensured by the service provider. Microsoft Sender ID [21] and Yahoo based Domain Key [22] provides some examples of domain level authentications. Email level authentication [23] is also used to authenticate email based on domain name and hash of password as digital signature. Most of the users do not use email authentication and that became one of the biggest drawback.

*c. Feature based email classification*

A very common strategy behind using phishing emails is to embed a link clicking on which leds the user to a fake webpage. The phishing email used email structure with embedded URI to ask the user to disclose confidential data. However, these phishing emails are extracted by using some features, which can easily detect by using previous knowledge as phisher repeat some pattern to fool their victims by disturbing the email feature.

i. Link features

A hyperlink structure is as : <a href="URI">"Visual Text"<\a>, where URI is the actual address of link of "Visual Text." URI contents are not displayed in web browser, but "Visual Text." Phisher uses this fact in 'bait' e-mails to redirect victim to a phishing websites. LinkGuard algorithm [24], examines the actual and visual link for any differences. It compares the actual DNS and the DNS visible, if they are not similar then website is definitely phishing. Use of IP address directly also signal that the website might be phishing, but it is not definite, if the destination information is missing actual DNS is examined, in case of encoded links, decoding is done followed by recursive execution of LinkGaurd.

ii. Structural features

Support Vector Machine (SVM) [25] which is deployed as server site to classify emails before they reach the client is the most commonly used classification mechanism for phishing emails. However, the experimental results in this work were not sufficient for large data, i.e. they used only 25 features to distinguish these email which were selected using simulated annealing.

iii. Word list features

This method is used to filters phishing emails that is applied either certain part or the whole email, the input here is a group of words without any sequence. The main approach of the classification is based on the machine-learning algorithm.

This model has various shortcomings, it requires a large number of features, it has high complexity in terms of time and memory, it is also not able to detect zero-day attacks.

i) k-Nearest Neighbour (k-NN) - Gansterer proposed k-NN classifier [26] classifies phishing emails based on k-nearest training input where training data is chosen using a predefined similarity function.

ii) Naive bays classifiers – it uses Bayes theorem to perform probabilistic classification, it is mostly used for text classification and keyword filtering. The features used for Naïve Bayes classifier are statistically independent to maintain the accuracy.

*d. Comparison of existing solutions*

Comparisons of various existing solutions for Phishing Website detection is shown in Table II.

Table II. Comparisons of various existing solutions for Phishing emails

| Reference | Solution | Utility | Approach | Limitation/ Remarks |
|---|---|---|---|---|
| DNSBL Information [20] | DNS-based blacklist | In anti-spam filters | Blacklist a range of IP address and domain names | Zero-day phishing |
| Lyon et. al., 2006 [21] | Sender ID | In Microsoft Sender ID | Domain level authentication is used by sending Sender ID | Both side must use the same technology |
| Delany et. al., 2007 [22] | Domain Key | In Yahoo Domain Key | Domain level authentication is used by sending Domain Key | Both side must use the same technology |
| B. Adida, et. al., 2005 [23] | Email authentication | Gmail, Hotmail, Yahoo | Authenticate by password hashing with domain name | Most user do not use email authentication |
| J. Chen, et. al., 2006 [24] | LinkGuard algorithm | In Windows it check all mailing application | Find the similarities between domain of actual and visual links | Check only for emails and higher false positive rate |
| M. Chandrasekaran, et. al., 2006 [25] | SVM based structural properties | Check emails before it reach to inbox | Implemented between MTA and MUA using SVM classifier | Use a very small set of features (25 only) |
| W. N. Gansterer, et. al., 2009 [26] | k-Nearest Neighbour | Rank emails in Ham, Spam and Phishing | Detect emails based on similarities in k-sample phishing emails | High false positive rate then the spam filters |
| Fette, Sadeh et al. [98] | PILFERS | Uses 10 features including Spam Assasin output. | Random forest and Support vector machine (SVMs) for classification | Large amount of emails are misclassified. |
| Bazarganigilani [42] | Based ontology concept and a set of heuristics | | The working is divided in 5 steps and uses Information Gain (IG), Naïve Bayes classifier | Low accuracy of classification. |
| Bergholz et al.[14] | Uses Dynamic Markov Models | Study statistical filtering of the phishing Emails | Dynamic markov chains are used to train the classifier on feature set. | High time and storage complexity. |
| Kumarguru et. al [79-81] | User Education based Approach | Assess the effectiveness of training materials , oniline tests, embedded training approaches etc., | Use of short training materials will enable users to read, immediate training after person becomes a phishing victim, | Participants were more educated than the average Intenet users. |
| MP-Shield [97] | Blacklist and Data minig based approach | Detects phishing activites in android based devices | Uses google balcklist API to extract attributes from network traffic and then performs classification | There is no method for the updation of the model to increase it's knowledge |
| Park et. al [98] | Based ontology concept | the syntactic similarity for sentences, and the subject and object of verb comparison | to determine the hidden intention of email from the computer perspective so that machines could more accurately detect phishing emails | High complexity |
| Tayal et. al [99] | Data minig based approach | Particle Swarm Optimization trained Classification Association Rule Mining | a new rule pruning scheme in order to reduce the number of rules and increasing the generalization aspect of the classifier | High time and storage complexity |
| PhishWHO [100] | difference between the target and actual identities of a webpage | Phishing webpage detection via identity keywords extraction and target domain name finder | Exploit URL patterns based on the proposed N-gram model to extract identity keywords | Cannot address visual cloning |

### *C. Protection from Phishing Websites*

A fake website seems to be similar with any legitimate websites in look and design. The URL pattern of any fake page also seems to be similar in first glance. Phishers try their best in both look and feel as well as in URL pattern so that more and more victim attract toward their fake page without knowing that they are under phishing attacks. As whole Internet, cannot be control from server side, so solutions available to detect these fake pages are only for the client side. Based on these characteristics, we classified many solutions for phishing website [16-18], under the following categories:

#### *a. Blacklist and Whitelist*

Blacklists consists of a list phishing URLs and IP addresses detected in the past, which are updated in certain intervals, whereas, whitelist is collection legitimate addresses and URLs. They do not provide security against zero-day attacks, as new address or site cannot be detected by these blacklists. Whitelists generally used to reduce 'false positive' rates. However, blacklist has lower FP rates then heuristic.

i. Google Safe Browsing API

Google Safe Browsing API [27] allows the user application to verify if a given URL is blacklisted or not. Although the protocol is still experimental, various browsers including Google Chrome and Mozilla Firefox use it. Google provides the current implementation of the protocol, and which has two blacklists namely ‘goog-phish-shavar’ (phishing) and ‘goog-malware-shavar’ (malware).

Google Safe Browsing API allows the client side applications to check if a URL is blacklisted from a list which is continuously updated by Google. Although the protocol is still experimental, various browsers use it. The list maintained at the client side and is updated periodically; however, if URL is changed even a little bit from the blacklisted URL would result in no match.

ii. PhishNet: Predictive Blacklisting

PhishNet [28] solves the problem of exact matching (if a URL is slightly changed version of blacklisted one then it remains undetected). It generates almost all possible variants of a URL using five different variation heuristics, which are:

i) Replace Top Level Domains (TLD)
ii) Directory structure similarity
iii) IP address equivalence
iv) Query string substitution
v) Brand name equivalence

iii. Automated Individual White-List

Automated Individual White-List (AIWL) [29] keeps a list of legitimate websites where users have given their sensitive information which are called trusted Login User Interfaces (LUIs) features. A warning is generated if credentials are submitted to any untrusted webpages not on the whitelist. Ye et al., proposed an approach in which a feature vector for successful and failed login attempts based on the AIWL is given to Naïve Bayes classifier which construct a model, which calculates probabilities of future attempts of login. The attempt is predicted to be successful if the probability is higher than a predefined threshold.

b. *Heuristic Solutions*

Heuristics refer to set of rules based on previous results and experiences, to solve a problem or learning purposes. The solutions based on heuristics are not necessarily optimal, but the results are near optimal and ease the decision making process.

Phishing detection based on heuristics are found to be effective in case of zero-day phishing attacks. These are based on the fact gathered from real time phishing attacks; however, it has the risk of high false positives, but gives better results than blacklist based approaches. However, Browsers such as, Mozilla Firefox, Internet Explorer etc., use heuristic based solution for phishing detection.

i. SpoofGuard

SpoofGuard [30] is browser plug for Internet Explorer in developed by Stanford university. It uses a set of heuristics to detect anomalies in the webpage content. It detects phishing scams based on HTTP. It defines a certain threshold value and if the results of the heuristics cross the threshold level a warning is given to the user. SpoofGuard checks if the URL is similar to a whitelisted one. Then it detects for the presence of a hidden attribute in the URL. If the URL in the text attribute is different from the actual one then the site malicious. It has traffic indicator system, which calculates the threat level by navigating the site. The passwords fields if present also increase the level of threat (though they are mostly harmless) as they might copy a login form.

ii. Collaborative Intrusion Detection

In CIDS [31], untrusted data is exchanged between different Intrusion Detection Systems (IDSs). Each CIDS examines DNS cache to retrieve high resource records zones and low Time-to-live values. This list is sent to a global CIDS. All the systems then monitor the addresses on that list and detect the infected ones. The detection of origin of the malicious content can be done by analyzing incoming and outgoing connection of suspicious IP address. But this approach is not implemented due to complexity regarding examining the fast flux attacks.

iii. PhishGuard: A Browser Plug-in

Phishguard [32] is a heuristic based approach, which performs phishing detection based on HTTP authentications. It is based on the fact that phishing sites only store the credentials for future use and do not verify them. PhishGuard starts it test when web page requires user credentials. It sends the same user ID but a different password to the pages, if the response is HTTP 200 then the page is phishing. In case of HTTP 401 response, either it is wrong password error or the website is blindly signaling failed authentication. Although there is a fair chance of the site being legitimate, the final decision is made based on the hash of the password. The site is regarded as phishing if it already has hash value, else it is legitimate and the user is requested to re-enter the password.

iv. CANTINA: A Content-Based Approach

CANTINA [33] is a toolbar that examines webpage content for phishing detection by calculating Term Frequency-Inverse Document Frequency (TF-IDF) for each term on the webpage, the top n terms with highest values are then used to represent that document. They are given as search query in of a search engine, and domains of the first n entries are stored. If the webpage is one of these entries then it is considered to be legitimate else not. The following set of heuristics is used to lower the false positives:

(i) A domain more than 12 months old is likely to be legitimate.
(ii) Presence of – or @ in the linkor URL indicates page is phishing.
(iii) Presence of more than five '.' Indicates page is phishing.
(iv) Embedded HTML forms indicate page is phishing.

c. *Visual Similarities*

One of the important properties, which is maintained by almost every phishers, so that the victim cannot easily distinguish between a fake page and a trusted target page. If any fake page is not similar in visual appearance then there is very less chance to make fool any victim easily. Visual appearance is one of the things, which any person noted first compare to any other things like URI or SSL certification or https protocol in address bar. Based on these properties, researcher also proposed their solutions as given below:

i. Visual Similarity Based Phishing Detection

Visual Similarity Based Phishing Detection (VSBPD) [34] detects when a user is giving any information to an untrusted webpage. It keeps a check on the forms filled by the user, it looks for the similarities of text and images embedded on the page. It also stores user credentials and where they are to be sent. If the website is not on the trusted list, the processes is interrupted and a warning is generated. The warning is generally raised when there is similarity between two pages, in case both of them require same information, however it is less likely for any of these websites to be fake if they not similar in appearance. This approach is inspired by Anti-Phish (plug-in) and DOMAntiPhish (browser extension).

ii. BaitAlarm

According to [35], the increase in similarity between phishing and legitimate page increase the chance of user falling for that phishing scam. Thus, phisher always try to make fake pages that have very subtle differences from the target page. They use CSS technology to maintain the consistency of the page.

The BaitAlarm consists of three components: Pre-Processor, Layout Monitor, and Network Library. Pre-Processor extract the layout of any new loaded page, and when user enter credentials in this page, browser hold this page and the layout information is sent to Layout Monitor. When Layout Monitor gets these information then comparison-unit extract CSS from page and check Network Library (maintain user history and table of victim pages) for the victim pages comparison-unit. Similarity score is calculated between targets and suspect page, and if score is found less than present threshold, the page is innocent, otherwise found a phishing page.

*d. Miscellaneous Solution*

There are various solutions present, which cannot classify among any of the above categories. These solutions can categories in miscellaneous solutions, as these solutions are not recent. However, these solutions have the historical impact as they came with evolution in the Internet technologies. Now a day, these solutions either embedded within the browser itself or used by various financial websites to overcome cyber-crime's problems. These solutions are described below:

i. TrustBar

TrustBar [36] monitors the top portion of browser window that has logos and graphical icons. It must be present for every window in the browser so that attacks in which a fake site hide browser indicators or exchange them with other indicators are detected and prevented.

ii. Dynamic Security Skin

It is an extension for Mozilla browser. It requires Secure Remote Password (SRP) protocol to authenticate Webpages. This extension has a trusted window for entering username and password; it deploys images to create a reliable path between the window and user to prevent any fake webpage and text field entries [37].

*e. Comparison of existing solutions*

Comparisons of various existing solutions for Phishing Website detection is shown in Table III.

Table III. Comparison of existing Phishing website solutions

| Reference | Solution | Utility | Approach | Limitation/Remark |
|---|---|---|---|---|
| Google Developer [27] | Google safe browsing API | Chrome, Firefox, etc. | Provide a blacklist and when any hit occur, browser block the page | Not able to detect Zero-day phishing and when IP change |
| P. Prakash, et. al. (2010) [28] | PhishNet | Predictive blacklist | Remove the exact match limitation of blacklist by various techniques | Not able to detect Zero-day phishing |
| W. Han, et. al. (2012) [29] | AIWL | Tool to maintain individual whitelist | User maintained their own individual whitelist & list of features of these legitimate webpages | AIWL warn whenever any information is sent to any other page then in the list |
| N. Chou, et. al. (2004) [30] | SpoofGuard | Internet Explorer plug-in | It detect spoofed pages based on URIs with the help of certain rules | If URIs are not as defined in rules, it cannot detect it |
| Y. Wu, et. al., (2003) [31] | CIDS | Intrusion Detection System | Exchange data among IDS globally | Not implemented yet |
| Y. Joshi, et. al. (2008) [32] | PhishGuard | A browser plug-in | A phishing website don't respond correctly while asking of credentials | Credentials theft if phishing website reply unauthorised |
| Y. Zhang, et. al. (2007) [33] | CANTINA | Internet Explorer toolbar | Search top TF-IDF in search engine and find current URI in top list | Higher false positive rate when TF of any other term is high |
| E. Medvet, et. al. (2008) [34] | Visual Similarity based Phishing Detection | Approach to detect phishing | Find similarities based on text pieces, image embedded and overall visual appearance | Not distinguish if text pieces are replace with image of same appearance |
| J. Mao, et. al. (2013) [35] | BaitAlarm | Google Chrome | Compare CSS of two websites, where first is victim and second is phished | Selection of victim site is manually, which is not feasible in practical |
| Herzberg, et. al. (2004) [36] | TrustBar | Secure user interface add-on to browsers | Identify SSL certificate and shown in browser | Implemented in browser itself and not used separately |
| R. Dhamija, et. al. (2005) [37] | Dynamic Security Skin | Extension for Mozilla Browser | Generate unique 'skin' for each user and each transaction | Depend on human, whether or not he/she understand the 'skin' |
| Chen, et. al. (2010) [38] | Normalized Compression Distance | Approach to detect phishing Website attacks | Used the concept of Gestalt theory and super signals to treat webpages as indivisible | Not robust when alter to web pages e.g. image colour, relocation of objects and text content |
| Gastellier - Prevost, et. al. (2011) [39] | Phishark | Anti-phishing toolbar | Define 20 heuristics for detecting phishing webpages | Not robust when alter to web pages and need to add more heuristics for detecting phishing webpages |
| Moghimi et. al (2016) [101] | Rule-based phishing detection method | PhishDetector extension | two feature sets to determine the webpage identity, a rule-based method by extracting the hidden knowledge from our model | not entirely reliable in detecting phishing attacks |
| Solanki et, al. (2016) [102] | Heuristic Based Approach | Approach to detect phishing Website attacks | extract the features then apply this features to machine learning techniques it will identify website are phished or legitimate | heuristic evaluation on an interface is minimal due to organizational constraints. |

# V. PHISHING AND INTERNET OF THINGS

The IoT architecture which has been predicted since last few years is now changing our old lives fundamentally by being a part of our daily tasks. Devices are made to be smart and connected to the Internet. IoT is everywhere from our homes, schools vehicles to our bodies [73]. Although it provides a radical prospect of making our lives more comfortable, the security professionals are trying to come up with solutions to cope up with the threats these devices are vulnerable against. These threats are far wider in the IoT scenario and with its growth the threats will grow too. In a highly connected environment the security threats can target organization, governments or common people and can result privacy exploitation or data thefts and so on [52]. SANS institute reports stated that in first half of 2015 Phishing attacks resulted into 37% intrusions in IT organizations [53]. The attackers basically use emails to lure the victims into falling for these attacks. These emails might have links to some malicious webpages; in general, these are available only for few hours, thus reducing the significance of blacklists and other heuristic approaches.

In addition to these types of attacks and their defenses in the current scenario some newest areas also require protection from phishing attacks and IoT is one of them. The IoT enabled gadgets and sensors have become another easy medium for the hackers. The attacks on IoT gadgets started to make headlines in January 2015, the security provider Proofpoint revealed a security attacks in which spam emails were sent three times a day in bulk and 25% devices hosting them were televisions and refrigerators alongwith routers [74]. This may be a wake-up call for the sectors which are based on developing IoT thingbots. In these cases, we cannot blame the user for ignorance as a phishing attack can be successful only if all the eight layers of security have been compromised, thus we need to make sure that these attacks get blocked at the initial levels. In the IoT the security is so weak that the hacker are able to use the software in the thingbots for relaying malicious emails without even sending a virus or Trojan, such devices can be easily used for DDos attacks without the user knowing it as it will not at all effect the devices functionality. And the only way to make these devices infection free is to take them offline regularly and update their software [75].

The IoT devices also are need are needed to be brought offline from time to time and updated as it the only way to disconnect it if it has become a part of botnet etc. The upgradation of devices can also be done from a remote server but it needs to be secured from intrusion. In a wireless sensor network the upgradation of nodes requires a secure authentication protocol to check each update received and detect any malware code [76]. In the year 2013, 20 billion devices were connected to Internet and this number will increase to 32 billion by the year 2020. Smart thing are the future and everyone is appreciating it but these devices are also making the job of attackers easy. Proofpoint showed in survey that during two weeks more than 100, 000 gadgets were compromised to send over 750,000 malicious emails [50].

Today an object mainly communicates with another object who is in the same application system, but there's no doubt that the technical future is connecting every application system and with the growth of the Internet of Things the communication between different systems will become more and more frequent for the collaboration. But as the lack of global standards, they may have used different standards and technologies, so the interoperability is a problem. Only if we can solve the interoperability problem we can have the Internet of Things with better connectivity. Some researchers [73] have come up with a solution that is addition of a Coordination Layer into the Internet of Things' architecture design. The coordination layer responses to process the structure of packages from different application systems and reassemble them to a unified structure which can be identified and processed by every application system. Of course, if the standards of the Internet of Things are completed then the systems which based on the standards will have no problem in interoperability.

To secure the IoT devices from such attacks we have designed a simple algorithm for detection of such email traffic to secure the household devices. The purpose of spam emails is mainly advertising whereas phishing emails always have some criminal intent. Sometimes even legitimate emails are classified as spams. Consider a situation that a user is browsing through his/her spam folder and encounters an email that seems to be fine but is a phishing email, and being unaware he/she ends up giving his/her credentials or installing a malware. This gives rise to a need of filtering phishing emails so that the user is warned not to open those mails.

Many security approaches have been proposed to secure IoT environment, but there are no such approaches for the detection of spam and phishing emails. Most of the approaches use encryption. But, these devices are not protected by anti-spam or anti-virus software, thus the IoT attacks cannot be resolved at the source site. Thus, the attacks on these devices are highly distributed in nature which results in malicious emails successfully reaching inboxes [50, 76].

## VI. CURRENT ISSUES AND CHALLENGES

In the literature survey, various solutions to control phishing attacks have been given however, we found that there is no solution which we can say 'bullet of silver' against phishing. With time, phishing is becoming a more common measure to commit e-crime. Every time, when researchers come up with any idea to detect and prevent phishing, phishers change their attack strategy by exploiting vulnerabilities found in the current solution. Therefore, we can say that, it is a race condition between phishers and researchers.

We have already discussed that the phishing scams can be run either by malware or social engineering which refers to the use of either fake web pages or emails [54, 55]. Thus, there are many solutions are available to detect these emails and websites. To control phishing emails successfully, there were various solutions proposed as discussed in [19-26]. Similarly, to control phishing websites attacks, there were also various solutions proposed in [16-18, 27-39]. Phishing email is not the only way to fraud, but phishers also uses fake websites and phishing emails have also the link of fake websites. Phishing websites found more harmful as compare to phishing email, as many of phishing emails are filter before receiving by user inbox and filtered mail found under the spam mail. Any educated user can easily neglect these filtered spam mail if filtered correctly.

DNS-based Blacklist (DNSBL) [20] deploy the DNS protocol to control phishing emails. But due to large number of blacklisted a server faces constraints in terms of resource and performance, if it is not optimal for handling large number of DNS records. The blacklists are required to be updated periodically which requires an interactive behavior and the attackers take advantage of it if they have access to a legitimate PC or by changing IP addresses. In addition, it is unable to prevent from 'zero-day' phishing, i.e. initial victims cannot protect from phishing.

Spam Filtering techniques used at server side are also not very effective in case of phishing emails, as they perform classification on the appearance of certain words or phrases, if phisher changes in its statistics of phishing attacks, i.e. if there is solution present in server-side, which based on bag-of-words, then phisher have to avoid only these words and then rest is easy as 'piece-of-cake'.

User and server authentication approaches check whether the attacker is not pretending to be a valid sender of an email or a resource request, it increases the security at both server and user level. At user level authentication is ensured by use of passwords, but it is evident in past that passwords can be cracked by the phishers [21]. Authentication at domain level [22] is ensured by the service provider. Email level authentication [23], is also used to authenticate email based on domain name and hash of password as digital signature. Most of the users do not use email authentication and that became one of the biggest drawback.

Spoofed hyperlinks in the phishing mails are very common feature. LinkGuard algorithm [24], examines the actual and visual link for any differences. Support Vector Machine (SVM) [25] which is deployed as server site to classify emails before they reach the client is the most commonly used classification mechanism for phishing emails. However, the experimental results in this work were not sufficient for large data, i.e. they used only 25 features to distinguish these email.

Google Safe Browsing API [27] allows the client side applications to check if a URL is blacklisted from a list which is continuously updated by Google. Although the protocol is still experimental, various browsers use it. The list maintained at the client side and is updated periodically; however, if URL is changed even a little bit from the blacklisted URL would result in no match. PhishNet [28] addresses the exact match limitation found in blacklists. As life of these, phishing attacks are very less, a large amount of data is consumed to store these blacklisted URIs and domain, which have no use in near future. In addition, the complexity of comparing every URI with blacklist data is very high. Most common vulnerability of blacklist schemes are that the security still compromises as phisher still run the site by just changing the IP address or by using bots to spoof the domain.

After blacklist scheme, some heuristic scheme proposed to detect phishing. Unlike blacklist detection technique, heuristic techniques can identify 'zero-day' attacks. But these schemes have large number of 'false positive' then blacklist schemes. Due to more advanced and complex scams it is difficult to design heuristics without false positives. These scheme lesser number of updates as compared to blacklisting and whitelisting approaches but they have high time complexity.

Visual based similarity, on the other hand is appropriate scheme to detect any website as phishing. To do so, phisher always use visual similarity with the target website. Visual Similarity Based Phishing Detection (VSBPD) [34] gives a warning to user whenever he tries gives his credentials to an untrusted website. It checks the visual appearance of a page including images and font etc., and also remembers the details user is giving to a page and where it is to be sent.

BaitAlarm [35] is comparatively more efficient as VSBPD compare the text and their style in two websites, but if text content is replaced with picture then this scheme cannot able to compare these pages. BaitAlarm on the other hand compare the CSS of two websites, though the phishing site give same look yet they have very subtle differences with respect to the content, this approach uses visual similarities for phishing website detection. However, both scheme not describe that, how we choose a legitimate site from which, we compare any suspicious site. Storing information that describes image can be expensive as image take more space then data of any page used in heuristic schemes. In addition, large 'false positive' found as in heuristic schemes.

In the recent years, Phishing attacks have become one of the most serious threats faced by the Internet users, organizations and service providers. Several approaches have been proposed in the literature for the detection and filtering of phishing attacks, however Internet community is still looking for a complete solution to secure the Internet from these attacks. The same is true for the lightweight IoT devices, where the botnets are already a major concern. These are virtually no ways to detect any breaches in these devices but to take them offline and update their software manually.
Many security approaches have been proposed to secure IoT environment, but there are no such approaches for the detection of spam and phishing emails. Most of the approaches use encryption. But, these devices are not protected by anti-spam or anti-virus software, thus the IoT attacks cannot be resolved at the source site. Thus, the attacks on these devices are highly distributed in nature which results in malicious emails successfully reaching inboxes.

## VII. CONCLUSION AND SCOPE FOR FUTURE WORK

It has been a couple of decades since the phishing problem arose, but it is still used to steal personal information, online credentials, and credit card details. There are various solutions available, but whenever any solution proposed to overcome these attacks, phishers came with the vulnerabilities of the solution to make the attach successful. From these attacks, we focus on the social engineering attacks, as it creates negative effect on online commerce. Phishers always used communication media for the fraudulent activities using spoofed emails and fake websites. It creates bad impression on e-commerce, which is very much necessary in this new era of Internet. Our survey helps new researchers to understand the history, current trends of attacks and failure of various available solutions.

We classified social engineering phishing based on spoofed email attacks and fake websites. We have also classified various solutions either in the spoofed email filtering or in fake page detection. We further classify these solutions as per some common properties share between them. These classifications are based on blacklist, network, heuristics, some feature and various other properties. After the classification, we also described various issues and challenges in current solutions to understand the idea for future study to help the humanity by defending against phishing attacks.